\documentclass[10pt,letter]{article}

\usepackage{authblk}
\usepackage{amsmath}
\numberwithin{equation}{section}
\usepackage{amssymb}
\usepackage[table,xcdraw]{xcolor}
\usepackage{cancel}
\usepackage[normalem]{ulem}
\usepackage{graphicx}
\usepackage{setspace}
\usepackage{fullpage}  
\usepackage{amsthm}

\newcommand\etal{\textit{et al.}}
\newcommand{\defn}{\textit}
\newcommand{\Ord}{\mathrm{O}}
\newcommand{\alphac}{\alpha_\mathbf{c}}
\newcommand{\alphacz}{\alpha_\mathbf{c}^{(0)}}

\title{Improved estimates for the number of non-negative integer matrices\\
with given row and column sums}
\author[1]{Maximilian Jerdee}
\author[2,3,4]{Alec Kirkley}
\author[1,5]{M. E. J. Newman}

\affil[1]{Department of Physics, University of Michigan, Ann Arbor, MI 48109.  U.S.A.}
\affil[2]{Institute of Data Science, University of Hong Kong, Hong Kong}
\affil[3]{Department of Urban Planning and Design, University of Hong Kong, Hong Kong}
\affil[4]{Urban Systems Institute, University of Hong Kong, Hong Kong}
\affil[5]{Center for the Study of Complex Systems, University of Michigan, Ann Arbor, MI 48109.  U.S.A.}
\begin{document}
\maketitle

\begin{abstract}
  The number of non-negative integer matrices with given row and column sums appears in a variety of problems in mathematics and statistics but no closed-form expression for it is known, so we rely on approximations of various kinds.  Here we describe a new such approximation, motivated by consideration of the statistics of matrices with non-integer numbers of columns.  This estimate can be evaluated in time linear in the size of the matrix and returns results of accuracy as good as or better than existing linear-time approximations across a wide range of settings.  We also use this new estimate as the starting point for an improved numerical method for either counting or sampling matrices using sequential importance sampling.  Code implementing our methods is provided.
\end{abstract}

\section{Introduction}
Matrices with non-negative integer elements and prescribed row and column sums arise in a range of statistical, physical, and mathematical contexts.  For example, they appear in statistics and information theory as \defn{contingency tables}, whose elements count the number of times a state or event~A occurred, contingent on the occurrence of another state or event~B.

An important but difficult problem is to compute the number of matrices for given values of the row and column sums, i.e.,~the number $\Omega(\mathbf{r},\mathbf{c})$ of $m \times n$ non-negative integer matrices whose rows sum to $\mathbf{r} = (r_1,\ldots,r_m)$ and whose columns sum to $\mathbf{c} = (c_1,\ldots,c_n)$.  This number plays a role for instance in the calculation of mutual information measures for classification and community detection~\cite{newman2020improved} and in sequential importance sampling methods for integer matrices~\cite{chen2005sequential,harrison2013importance}.  

No exact analytic expression is known for~$\Omega(\mathbf{r},\mathbf{c})$ for general $\mathbf{r}$ and~$\mathbf{c}$, and evaluating it exactly by numerical means is known to be \#P-hard~\cite{dyer1997sampling}, meaning it is improbable that an algorithm exists with runtime polynomial in $m$ and~$n$ for general $m,n$.  Workable exact algorithms do exist for small~$m,n$~\cite{barvinok1994polynomial} and for cases with bounded row or column sums~\cite{miller2013exact}, but outside of these limited settings the only tractable approach is approximation.  In this paper we review previous approximation methods for this problem and present a new, computationally efficient approximation that is simple to implement and compares favorably with previous approaches in terms of both accuracy and running time.

Previous approximation methods for this problem fall into three broad classes, which we will refer to as \defn{linear-time}, \defn{maximum-entropy}, and \defn{sampling-based} methods.  The majority of the approaches fall into the first category of linear-time methods, which are characterized by their rapid $\Ord(m+n)$ computation time, although they achieve this efficiency at the expense of accuracy and scope.  The linear-time approaches include methods based on combinatoric arguments~\cite{good1976application,good1977enumeration} and moment-matching arguments~\cite{diaconis1985testing,gail1977counting}, and methods tailored to the sparse regime~\cite{bekessy1972asymptotic,greenhill2008asymptotic} in which most elements of the matrix are zero.  The method we propose also falls into the linear-time category and consistently performs near the top of this class across a wide array of test cases.

The second class of methods are maximum-entropy methods, developed in this context by Barvinok and Hartigan~\cite{barvinok2012matrices}. For large $m$ and~$n$ these methods outperform the linear-time methods in terms of accuracy but are much slower, requiring the numerical solution of a continuous convex optimization problem followed by evaluation of an $(m+n-1)\times(m+n-1)$ matrix determinant, for a time complexity of about $\Ord((m+n)^3)$.  The basic method employs a Gaussian maximum-entropy approximation but the result can be further refined using an ``Edgeworth correction,'' which requires an additional $\Ord(m^2 n^2)$ computation but substantially improves accuracy. 

The third class of approximations are sampling-based methods, including Markov-chain Monte Carlo (MCMC) methods~\cite{holmes1996uniform} and sequential importance sampling (SIS)~\cite{chen2005sequential}.  Given sufficient running time these methods are expected to converge to the true answer, although the time taken can be prohibitive.  SIS is typically better than MCMC for calculating~$\Omega(\mathbf{r},\mathbf{c})$ in terms of both speed and accuracy~\cite{chen2005sequential,harrison2013importance}, and we make use of the SIS method in this paper to establish benchmarks for the evaluation of the other methods.  As a bonus, the new linear-time approximation we propose can also be used to improve the convergence of SIS, allowing us to apply the latter method to substantially larger matrices than has previously been possible.

In addition to the specific problem of counting integer matrices with given row and column sums, a number of other related problems have received attention.  The problem of sampling such matrices uniformly arises in a variety of contexts and can be tackled efficiently by a modification of the SIS algorithm we propose in Section~\ref{SIS}.  The problem of counting matrices whose elements take the values 0 and 1 arises in graph theory and, although it is not our main concern here, our methods can be extended to this case also (with some caveats) and we compare the results with a variety of competing methods in Appendix~\ref{app:ZeroOneTables}.  Finally, there are certain special cases of the matrix counting problem, such as cases where the row and column sums are uniform (so-called magic squares), for which one can make significant progress beyond what is possible in the general case~\cite{rodney2010asymptotic}.  We will not discuss these special cases here however: our focus in this paper is on the general case.

\section{Summary of results}
\label{summary}
This paper presents a number of new results.  First, we derive a new and simple linear-time approximation for the number of non-negative integer matrices with given row and column sums $\mathbf{r},\mathbf{c}$ thus:
\begin{align}
\label{eq:ECEstimate0}
\Omega(\mathbf{r},\mathbf{c}) &\simeq \binom{N+m \alphac - 1}{m \alphac - 1}^{-1} \prod_{i=1}^m \binom{r_i + \alphac - 1}{ \alphac - 1} \prod_{j=1}^n \binom{c_j+m-1}{m - 1},
\end{align}
where
\begin{align}
\label{eq:alphaC}
\alphac &= \frac{N^2 - N + (N^2 - c^2)/m}{c^2 - N}, \\
N &= \sum_i r_i = \sum_j c_j, \qquad
c^2 = \sum_{j=1}^n c_j^2.
\end{align}
(There are certain trivial cases where Eq.~\eqref{eq:alphaC} gives invalid values for~$\alphac$ but these are easily dealt with---see Appendix~\ref{app:ECdetails}.)

\begin{figure}[t]
\centering
\includegraphics[width=16cm]{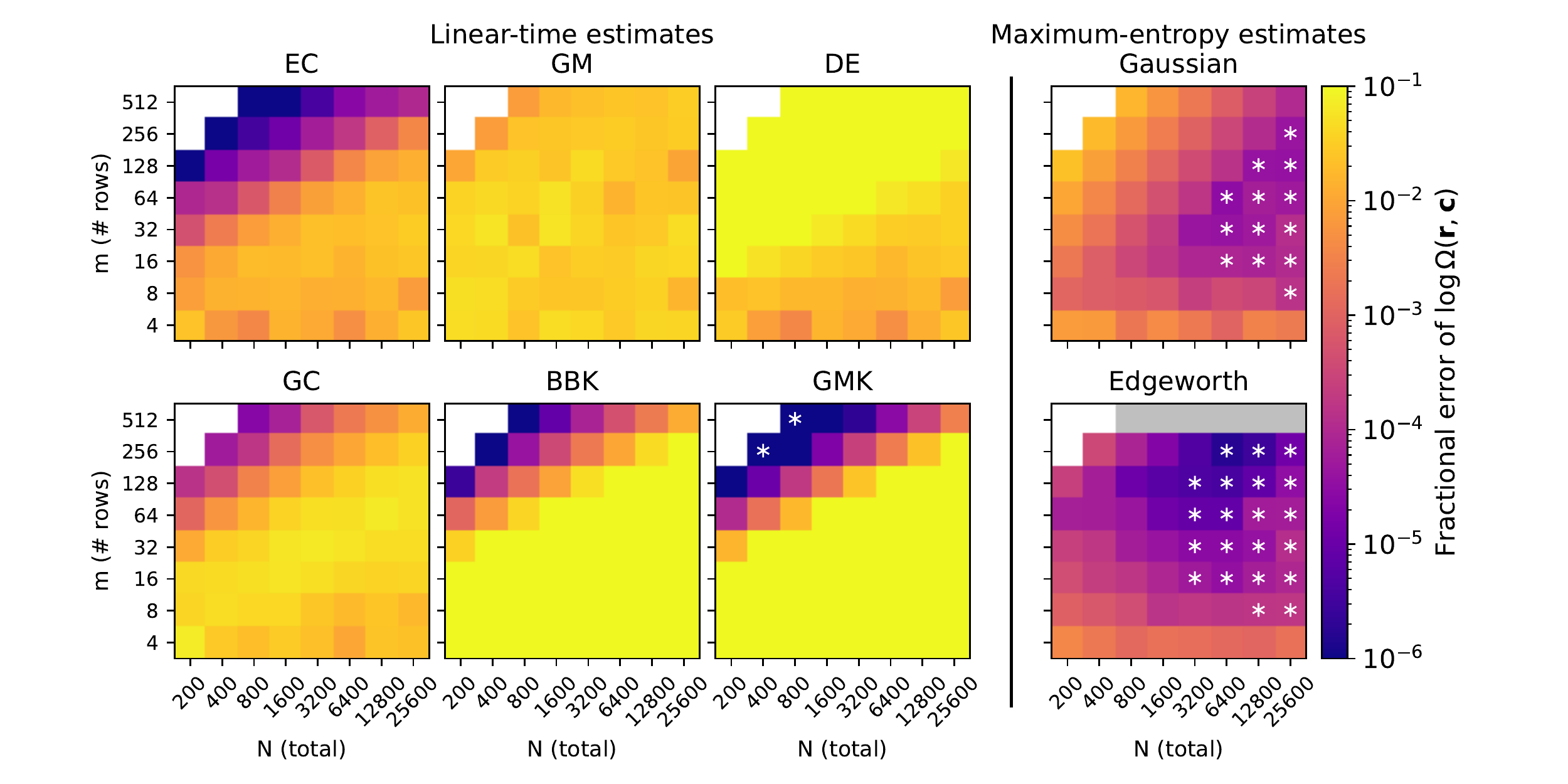}
\caption{Fractional error in various estimates of $\log \Omega (\mathbf{r},\mathbf{c})$ for square $m \times m$ matrices with total sum~$N$, relative to ground-truth results computed by sequential importance sampling (see Section~\ref{SIS}).  Each square represents an average over ten sets of margins $\mathbf{r},\mathbf{c}$ drawn uniformly at random.  Asterisks denote data points for which the error is within five times the estimated error from the sequential importance sampling, and so the true error may be smaller in these cases.  White squares indicate invalid parameter combinations where $m>N$ so that margins $\mathbf{r},\mathbf{c}$ can not be generated without zeros.  Gray regions indicate parameter values for which estimates could not be computed in an hour of run time or less.  See Appendix~\ref{app:numericalDetails} for further details of the benchmarking and related issues.}
\label{fig:Grid-Random}
\end{figure}

Second, we have conducted exhaustive tests of this estimate and five other previously published linear-time estimates, comparing them with ground-truth results derived from sequential importance sampling.  Figure~\ref{fig:Grid-Random} summarizes the results of our tests.  We find that most of the differences in performance can be revealed by considering square $m \times m$ matrices of various sizes, while varying the sum~$N$ of all entries.  In our calculations we generate ten random test cases for each parameter pair~$N,m$ with margins $\mathbf{r}$ and $\mathbf{c}$ drawn uniformly from the set of $m$-element positive integer vectors that sum to~$N$. We then perform a lengthy run of sequential importance sampling (SIS) on each sampled test case to establish a ground-truth estimate of the number of matrices. Armed with the SIS estimates, we apply each of our six linear-time estimators to the same test cases and compute the error on each one.  We report performance in terms of the fractional error in $\log \Omega(\mathbf{r},\mathbf{c})$, since the logarithm is simpler to deal with numerically and is also often the quantity of most interest~\cite{newman2020improved}.

The first panel of Fig.~\ref{fig:Grid-Random}, labelled ``EC'' (for ``effective columns''---see Section~\ref{ECEstimate}), shows the results for our new estimator, Eq.~\eqref{eq:ECEstimate0}.  The running time for all of the linear-time estimators is negligible, but as the figure shows their accuracy varies.  In particular, we distinguish a sparse regime where $N\ll mn$ so that most matrix elements are zero (up and to the left in the plots) and a dense regime where $N\gg mn$ so that most matrix elements are nonzero (down and to the right).  Some estimates, such as those labeled BBK and GMK, perform well in the sparse regime but poorly in the dense regime.  Others, such as DE, do the reverse.  The EC estimate of this paper, however, is comparable to or better than the others in both the sparse and dense regimes, while still being fast and simple to compute.  In the dense regime the fractional error is around $10^{-2}$ or~$10^{-3}$, becoming as good as $10^{-6}$ in the sparse regime.  The estimate denoted GC also gives acceptable performance in both sparse and dense regimes, but is not competitive with the EC estimate in these tests.

We have also performed tests using the two maximum-entropy estimates of~\cite{barvinok2012matrices} for a portion of the same test cases.  As the figure shows, these estimates perform well across both the sparse and dense regimes and in general outperform the linear-time estimates, including our own, but at the expense of much greater computational effort.  As mentioned in the introduction these estimates have time complexities of~$\Ord((m+n)^3)$ for the Gaussian approximation and $\Ord(m^2n^2)$ for the Edgeworth version. Thus, for a typical case with $m = 128$ and $N = 3200$ our implementations of the linear-time estimates run in under 2\,ms, the Gaussian maximum-entropy method takes 3~seconds, and the Edgeworth-corrected version takes 22~seconds.  For our largest test cases with $m=512$ the calculation of the Edgeworth correction becomes so demanding as to be impractical, so results for these cases are omitted from Fig.~\ref{fig:Grid-Random}.

\begin{figure}
\centering
\includegraphics[width=13cm]{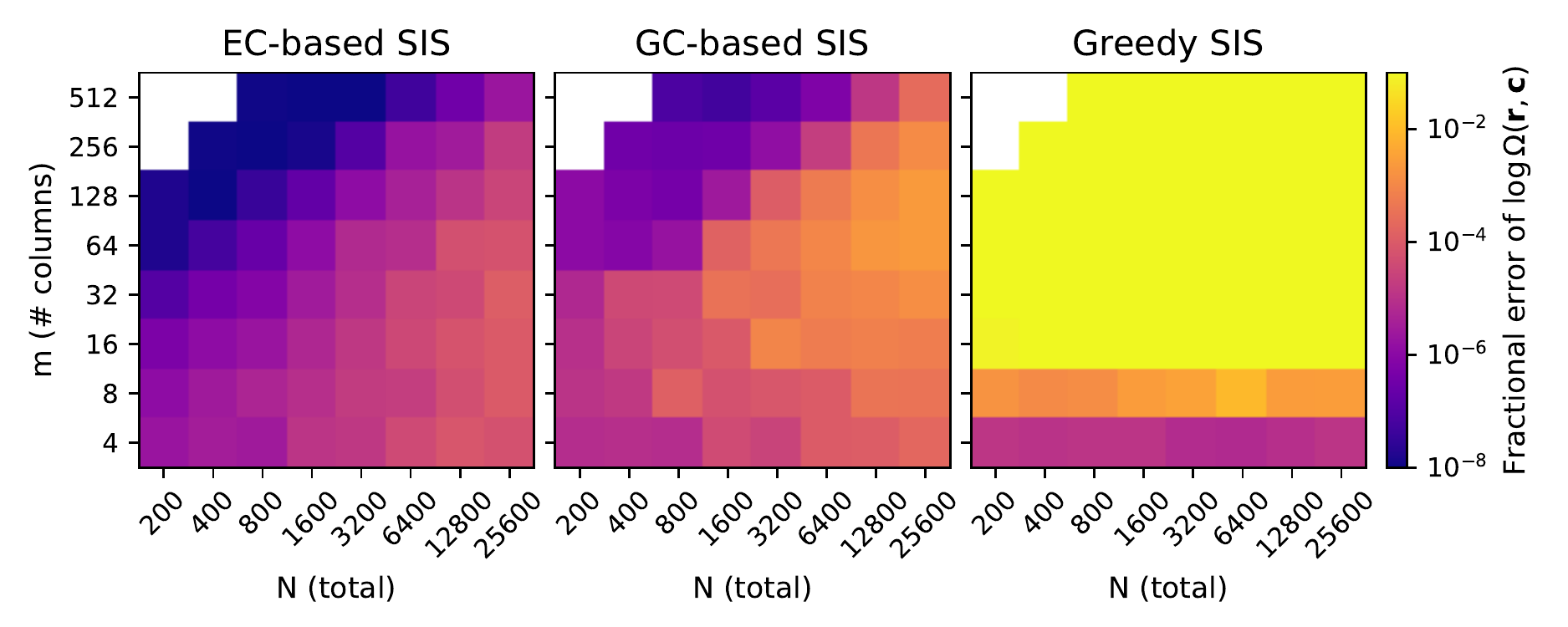}
\caption{Fractional error on sequential importance sampling estimates of $\log \Omega(\mathbf{r},\mathbf{c})$ for three different variants of the SIS approach. The EC-based SIS serves as the benchmark for the results in Fig.~\ref{fig:Grid-Random}. Note that the color scale in this figure differs from that in Fig.~\ref{fig:Grid-Random}.}
\label{fig:SIS-Comparison}
\end{figure}

Aside from their substantial computational demands, the maximum-entropy methods work particularly well in the regime of intermediate-to-high density and indeed do so well in this region that their accuracy becomes comparable to the accuracy of the sequential importance sampling that we use to compute the ground truth.  The SIS calculation, like all sampling methods, displays some statistical error, as shown in Fig.~\ref{fig:SIS-Comparison}.  Although this error is usually negligible, it is a limiting factor for evaluating the maximum-entropy estimates in some cases.  These cases are denoted by asterisks in Fig.~\ref{fig:Grid-Random}.

Software implementations of the various estimates and SIS methods described in this paper may be found at \verb|https://github.com/maxjerdee/contingency_count|. 

\section{Linear-time estimates}
\label{analytic}
Turning now to the details, in this section we discuss the linear-time methods for estimating~$\Omega(\mathbf{r},\mathbf{c})$.  We first present our new estimate, which is based on the concept of ``effective columns.''  We also describe three related approaches due to Gail and Mantel~\cite{gail1977counting}, Diaconis and Efron~\cite{diaconis1985testing}, and Good and Crook~\cite{good1976application,good1977enumeration} and two somewhat different approaches tuned to the sparse case and due to B\'ek\'essy, B\'ek\'essy, and Koml\'os~\cite{bekessy1972asymptotic} and Greenhill and McKay~\cite{greenhill2008asymptotic}.

\subsection{A new estimate for matrix counts}
\label{ECEstimate}
In this section we derive the approximation for $\Omega(\mathbf{r},\mathbf{c})$ given in Eq.~\eqref{eq:ECEstimate0}.  Let $A(\mathbf{c})$ be the set of all non-negative $m \times n$ integer matrices $X = (x_{ij})$  with column sums~$\mathbf{c}$ but unconstrained row sums:
\begin{align}
A(\mathbf{c}) \equiv \Big\{\{x_{ij}\} \in \mathbb{N}^{m \times n} \Big| \sum_{i} x_{ij} = c_j, j = 1,\ldots,n\Big\}.
\end{align}
By standard arguments the number of ways to choose the entries of column~$j$ so that they sum to~$c_j$ is $\binom{c_j + m - 1}{m - 1}$ and the columns are independent so the number of matrices in the set~$A(\mathbf{c})$ is
\begin{align}
|A(\mathbf{c})| = \prod_{j=1}^n \binom{c_j + m - 1}{m - 1}.
\label{eq:AcCount}
\end{align}
Now we further restrict to the subset of matrices $A(\mathbf{r},\mathbf{c}) \subseteq A(\mathbf{c})$ with both row and column sums fixed:
\begin{align}
A(\mathbf{r},\mathbf{c}) \equiv \Big\{\{x_{ij}\} \in \mathbb{N}^{m \times n}  \Big| \sum_{i=1}^m x_{ij} = c_j, \sum_{j=1}^n x_{ij} = r_i\Big\}.
\end{align}
The quantity we want to calculate is the number of matrices in this set $\Omega(\mathbf{r},\mathbf{c}) = |A(\mathbf{r},\mathbf{c})|$, in terms of which we can write the conditional probability $\mathbf{Pr}(\mathbf{r}|\mathbf{c})$ of observing a particular row sum~$\mathbf{r}$ given a uniform distribution over $A(\mathbf{c})$ as
\begin{align}
\mathbf{Pr}(\mathbf{r}|\mathbf{c}) = \frac{|A(\mathbf{r},\mathbf{c})|}{|A(\mathbf{c})|} = \frac{\Omega(\mathbf{r},\mathbf{c})}{|A(\mathbf{c})|}.
\end{align}
Since we have an exact expression for $|A(\mathbf{c})|$ in~\eqref{eq:AcCount}, the problem of calculating $\Omega(\mathbf{r},\mathbf{c}) = \mathbf{Pr}(\mathbf{r}|\mathbf{c})\, |A(\mathbf{c})|$ is thus reduced to one of finding~$\mathbf{Pr}(\mathbf{r}|\mathbf{c})$.  This is still a difficult problem and requires making an approximation.  Inspired by work of Gail and Mantel~\cite{gail1977counting} we take a variational approach and propose a family of candidate approximant distributions for~$\mathbf{Pr}(\mathbf{r}|\mathbf{c})$ then choose the best member of this family using a moment-matching argument.

Motivated by Diaconis and Efron~\cite{diaconis1985testing}, our family of candidate distributions is based on the unconditional distribution on the row sums~$\mathbf{r}$ (i.e.,~without any constraint on the column sums).  By the same argument that led to Eq.~\eqref{eq:AcCount}, the number of matrices with row sums~$\mathbf{r}$ is
\begin{align}
|A(\mathbf{r})| = \prod_{i=1}^m \binom{r_i + n - 1}{n - 1}.
\label{eq:ArCount}
\end{align}
At the same time the set~$A$ of all non-negative $m \times n$ integer matrices that sum to~$N$ has size
\begin{align}
|A| = \binom{N + mn - 1}{mn - 1},
\end{align}
and hence, under a uniform distribution over~$A$, the probability of observing row sum~$\mathbf{r}$ is
\begin{align}
\mathbf{Pr}(\mathbf{r}) = {|A(\mathbf{r})|\over|A|}
  = \binom{N + mn - 1}{mn - 1}^{-1} \prod_{i=1}^m \binom{r_i + n - 1}{n - 1}.
\end{align}
The key step in our argument involves approximating $\mathbf{Pr}(\mathbf{r}|\mathbf{c})$ by this unconditional distribution, but with the number of columns~$n$ replaced with a free parameter~$\alphac$, which we call the number of \defn{effective columns}:
\begin{align}
\mathbf{Pr}(\mathbf{r}|\mathbf{c}) \simeq
\mathbf{Pr}(\mathbf{r}|\alphac) = \binom{N + m\alphac - 1}{m\alphac - 1}^{-1} \prod_{i=1}^m \binom{r_i+\alphac - 1}{\alphac - 1}.
\label{eq:PrRGalpha}
\end{align}
The resulting distribution over $\mathbf{r}$ is the one that would be observed under the uniform distribution over $m \times \alphac$ matrices whose elements sum to~$N$.

Now we relax the constraint that~$\alphac$ be an integer, defining the obvious generalization of the binomial coefficient
\begin{equation}
\binom{n}{k} = {\Gamma(n+1)\over\Gamma(k+1)\Gamma(n-k)}.
\end{equation}
We are merely using~\eqref{eq:PrRGalpha} as trial distribution for our moment-matching argument, so the physical interpretation of $\alphac$ as an integer number of columns is not important.  As long as $\alphac>0$ the distribution is well-defined, normalized, and non-negative for every possible~$\mathbf{r}$.

In other contexts the distribution~\eqref{eq:PrRGalpha} is known as the (symmetric) Dirichlet-multinomial distribution.  When $\alphac=1$ the possible row sums~$\mathbf{r}$ are uniformly distributed among the possible non-negative integer choices of~$r_i$ that sum to~$N$.  As $\alphac\to\infty$ the distribution of $\mathbf{r}$ approaches a multinomial distribution where $\mathbf{r}$ is formed by taking $N$ samples from a uniform probability vector $(m^{-1},\ldots,m^{-1})$, the generalization of a symmetric binomial distribution.  For $0 < \alphac < 1$ the distribution of $\mathbf{r}$ will favor more extreme values of the coordinates~$r_i$, analogous to the behavior of the symmetric Dirichlet distribution.

Our approximation involves replacing the true distribution $\mathbf{Pr}(\mathbf{r}|\mathbf{c})$ by $\mathbf{Pr}(\mathbf{r}|\alphac)$, with the value of $\alphac$ chosen to make the approximation as good as possible in a certain sense.  To do this we use a moment-matching approach in which the value of $\alphac$ is chosen such that $\mathbf{Pr}(\mathbf{r}|\alphac)$ has the same mean and covariances as the true distribution $\mathbf{Pr}(\mathbf{r}|\mathbf{c})$.  Such a value always exists and it has a simple expression, as we now show.

The expectation and covariances of the $r_i$ under~$\mathbf{Pr}(\mathbf{r}|\alphac)$ are straightforward to compute:
\begin{align}
\mathbf{E}(r_i) = \frac{N}{m}, \qquad \mathbf{cov}(r_i,r_k) = (N/m) \frac{m\alphac+N}{m \alphac+1}(\delta_{ik} - m^{-1}).
\label{eq:ECEVar}
\end{align}
For $\mathbf{Pr}(\mathbf{r}|\mathbf{c})$ the calculation is only a little more involved.  In the uniform distribution over~$A(\mathbf{c})$ each column~$j$ is independently uniformly distributed over the possible choices of elements~$x_{ij}$ that satisfy $\sum_{i=1}^m x_{ij} = c_j$.  The expectations and covariances of these column entries alone are then
\begin{align}
\mathbf{E}(x_{ij}) = \frac{c_j}{m}, \qquad
\mathbf{cov}(x_{ij},x_{kj}) = \frac{c_j(m+c_j)}{m(m+1)}(\delta_{ik} - m^{-1}).
\end{align}
Since the true distribution $\mathbf{Pr}(\mathbf{r} | \mathbf{c})$ is the sum of these independent column distributions, the expectations and covariances add so that
\begin{align}
\mathbf{E}(r_{i}) = \sum_{j=1}^n \mathbf{E}(x_{ij}) = \frac{N}{m}, \qquad \mathbf{cov}(r_{i},r_{k}) = \sum_{j=1}^n \mathbf{cov}(x_{ij},x_{kj}) = \frac{N m+ c^2 }{m(m + 1)}(\delta_{ik} - m^{-1}),
\label{eq:rGcTrueEVar}
\end{align}
where we have introduced the shorthand $c^2 = \sum_{j=1}^n c_j^2$. 

Thus the expectations of $\mathbf{Pr}(\mathbf{r} | \alphac)$ and $\mathbf{Pr}(\mathbf{r} | \mathbf{c})$ already match and, equating the covariances in~\eqref{eq:ECEVar} and~\eqref{eq:rGcTrueEVar} and solving for~$\alphac$, we get
\begin{align}
\alphac = \frac{N^2 - N + (N^2 - c^2)/m}{c^2 - N}.
\label{eq:alphac}
\end{align}
Finally, we assemble (\ref{eq:AcCount}) and (\ref{eq:PrRGalpha}) into our ``effective columns'' estimate of $\Omega(\mathbf{r},\mathbf{c})$ thus:
\begin{align}
\Omega^{\text{EC}}(\mathbf{r},\mathbf{c}) =
      \text{Pr}(\mathbf{r}|\alphac)\,|A(\mathbf{c})|
    = \binom{N+m \alphac - 1}{m \alphac - 1}^{-1} \prod_{i=1}^m \binom{r_i+\alphac - 1}{\alphac - 1} \prod_{j=1}^n \binom{c_j+m-1}{m - 1},
\label{eq:ECEstimate}
\end{align}
where $\alphac$ is given by~\eqref{eq:alphac}.

Note that, although the number of matrices with given row and column sums is trivially symmetric under the interchange of rows and columns $\Omega(\mathbf{r},\mathbf{c}) = \Omega(\mathbf{c},\mathbf{r})$, our estimate of it is not.  (The same is also true of most of the other approximations discussed in this paper.)  The symmetry breaking occurs when we choose to approximate~$\mathbf{Pr}(\mathbf{r}|\mathbf{c})$ and not~$\mathbf{Pr}(\mathbf{c}|\mathbf{r})$.  In practice, as shown in Fig.~\ref{fig:Grid-Shape}, our estimate appears to perform better for matrices with more rows than columns, so it may improve performance to swap the definitions of $\mathbf{r}$ and $\mathbf{c}$ when $m<n$.  Conversely, if one requires a symmetric estimate, we recommend using the symmetrized form of $\log\Omega^{\text{EC}}(\mathbf{r},\mathbf{c})$ thus:
\begin{align}
    \log \Omega_{\text{S}}^{\text{EC}}(\mathbf{r},\mathbf{c}) = \tfrac{1}{2}\left[\log \Omega^{\text{EC}}(\mathbf{r},\mathbf{c})+\log \Omega^{\text{EC}}(\mathbf{c},\mathbf{r})\right].
\end{align}

\subsection{The estimate of Gail and Mantel}
\label{GMEstimate}
In the following sections we review some of the other approaches for estimating~$\Omega(\mathbf{r},\mathbf{c})$, outlining the motivation and derivations of the other linear-time approximations discussed in Section~\ref{summary}.  We start with an approach due to Gail and Mantel (GM)~\cite{gail1977counting}, who propose the following approximation for~$\Omega(\mathbf{r},\mathbf{c})$:
\begin{align}
\Omega^{\text{GM}}(\mathbf{r},\mathbf{c}) = \left(\frac{m-1}{2 \pi m \sigma^2}\right)^{(m-1)/2}m^{1/2} e^{-Q/2} \prod_{j=1}^n \binom{c_j+m-1}{m - 1},
\label{eq:GMEstimate}
\end{align}
where
\begin{align}
\sigma^2 = \frac{(c^2 + m N)(m-1)}{(m+1)m^2}, \qquad
Q = \frac{m-1}{\sigma^2 m} \left(r^2 - \frac{N^2}{m}\right), \qquad
r^2 = \sum_i r_i^2, \qquad
c^2 = \sum_j c_j^2.
\end{align}
The derivation of this estimate follows the same general logic as our own, with the true distribution $\textbf{Pr}(\mathbf{r}|\mathbf{c})$ approximated by a family of simpler distributions that is fitted to the true distribution with a moment-matching argument.  The difference lies in the particular family used: Gail and Mantel use a multinormal distribution, by contrast with the Dirichlet-multinomial distribution in our derivation.

Numerical results for the approximation of Gail and Mantel were given in Fig.~\ref{fig:Grid-Random}.  The results are only moderately good and the method is typically outperformed by the other linear-time estimators, indicating that there is some art to picking an appropriate family of distributions for the moment matching argument.  We note that in this case the multinormal distribution is not justified (as one might imagine) by the central limit theorem, despite $\mathbf{Pr}(\mathbf{r}|\mathbf{c})$ being a mixture of independent columns, because the probability density is typically evaluated away from the expected value of~$\mathbf{r}$, $\mathbf{E}(\mathbf{r}) = \left(N/m,\ldots,N/m\right)$, in a regime where central limit theorem arguments do not apply.

\subsection{The estimate of Diaconis and Efron}
\label{DEEstimate}
A related approximation has been put forward by Diaconis and Efron (DE)~\cite{diaconis1985testing}:
\begin{align}
\Omega^{\text{DE}}(\mathbf{r},\mathbf{c}) = \Bigl(N + \frac{mn}{2}\Bigr)^{(m-1)(n-1)}  \biggl(\prod_{i=1}^m \bar{r}_i\biggr)^{K_\mathbf{c} - 1}\biggl(\prod_{j=1}^n \bar{c}_j\biggr)^{m - 1} \frac{\Gamma(m K_\mathbf{c})}{\Gamma(m)^n\Gamma(K_\mathbf{c})^m}, \label{eq:DEEstimate}
\end{align}
where
\begin{align}
w = \frac{N}{N + \frac12 mn}, \qquad
\bar{r}_i = \frac{1 - w}{m} + \frac{w r_i}{N}, \qquad
\bar{c}_j = \frac{1 - w}{n} + \frac{w c_j}{N}, \qquad
K_\mathbf{c} = \frac{m+1}{m \bar{c}^2} - \frac{1}{m}, \qquad
\bar{c}^2 = \sum_j \bar{c}_j^2.
\end{align}
The derivation of this estimate follows similar lines to that for the estimates of this paper and of Gail and Mantel, Eqs.~\eqref{eq:ECEstimate} and~\eqref{eq:GMEstimate}, using a moment-matching argument, but with some crucial differences.  Instead of considering the set~$A(\mathbf{r},\mathbf{c})$ of integer matrices with the required row and column sums, Diaconis and Efron consider the space (polytope) $P(\mathbf{r},\mathbf{c})$ of all $m \times n$ matrices of non-negative reals (not just integers):
\begin{align}
P(\mathbf{r},\mathbf{c}) \equiv \Big\{\{x_{ij}\} \in \mathbb{R}^{m \times n} \Big| x_{ij} \geq 0 ,\sum_{i=1}^m x_{ij} = c_j, \sum_{j=1}^n x_{ij} = r_i\Big\}. \label{eq:Pdef}
\end{align}
The count $\Omega(\mathbf{r},\mathbf{c})$ of integer matrices can be thought of as the volume of the intersection of this polytope with the lattice formed by the (unconstrained) set~$A$ of non-negative integer matrices that sum to~$N$:
\begin{align}
\Omega(\mathbf{r},\mathbf{c}) =|A(\mathbf{r},\mathbf{c})| = |P(\mathbf{r},\mathbf{c}) \cap A|.
\label{eq:PIntersections}
\end{align}

Diaconis and Efron use moment-matching not to estimate $|A(\mathbf{r},\mathbf{c})|$ but instead to estimate the volume of the polytope  $P(\mathbf{r},\mathbf{c})$, then compute the size of the intersection~\eqref{eq:PIntersections} from it.  Since the polytope is a continuous region, the distribution $\mathbf{Pr}(\mathbf{r}|\mathbf{c})$ on it is also continuous and is represented with a continuous approximant distribution~$\mathbf{Pr}(\mathbf{r}|K_\mathbf{c})$, chosen to be $N$ times the symmetric Dirichlet distribution~$\mathop{\textrm{Dir}}(K_\mathbf{c})$, with the Dirichlet parameter~$K_\mathbf{c}$ chosen to match the mean and covariances of the true distribution~$\mathbf{Pr}(\mathbf{r}|\mathbf{c})$ over the polytope.  Armed with the resulting approximation for the volume of the polytope, $\Omega(\mathbf{r},\mathbf{c})$~is then estimated as the number of lattice points within it as in~\eqref{eq:PIntersections}, calculated from the volume of the polytope times the density of lattice points.  Finally, an ``edge-effects'' correction is applied to better reflect the number of lattice points contained.

For dense matrices the performance of this estimate is similar to that of our own estimate---see Fig.~\ref{fig:Grid-Random}.  Indeed, in the dense limit $N/mn \to\infty$ it can be shown that the two approaches are equivalent.  For sparse matrices, on the other hand, the approximation of the number of lattice points using the volume of the continuous polytope fails and the DE estimate breaks down.

\subsection{The estimate of Good and Crook}
Good and Crook (GC)~\cite{good1976application} proposed the following estimate for~$\Omega^{\text{GC}}(\mathbf{r},\mathbf{c})$:
\begin{align}
\Omega^{\text{GC}}(\mathbf{r},\mathbf{c}) &= \binom{N + m n - 1}{mn - 1}^{-1}\prod_{i=1}^m \binom{r_i + n - 1}{n - 1} \prod_{j=1}^n \binom{c_j + m - 1}{m - 1},
\label{eq:GCEstimate}
\end{align}
which is equivalent to our own estimate if one does not apply moment matching but instead simply assumes that the number of effective columns is equal to the number of actual columns: $\alphac = n$.  Under the circumstances, it seems likely that this estimate will be less good than that of Eq.~\eqref{eq:ECEstimate}, as indeed can be seen in the numerical results of Fig.~\ref{fig:Grid-Random}.

\subsection{Estimates for the sparse regime}
\label{Sparse}
The remaining two linear-time approximations presented in Fig.~\ref{fig:Grid-Random} are closely related and both aimed at approximating $\Omega(\mathbf{r}, \mathbf{c})$ in the sparse regime where $N \ll m n$ and most matrix elements are zero.  In this regime B\'ek\'essy, B\'ek\'essy, and Koml\'os (BBK)~\cite{bekessy1972asymptotic} give the following approximation:
\begin{align}
\Omega^{\text{BBK}}(\mathbf{r}, \mathbf{c}) &= \frac{N!}{\prod_{i=1}^m r_i! \prod_{j=1}^m c_j!}\exp\left[\frac{2}{N^2} \sum_{i=1}^m \binom{r_i}{2} \sum_{j=1}^n \binom{c_j}{2}\right]\,\bigl[1 + \Ord\bigl(\log N/\sqrt{N}\bigr)\bigr], 
\label{eq:BBKEstimate}
\end{align}
where the error term describes the asymptotic growth of the error with $N$ in the sparse limit where $\max(\mathbf{r})$ and $\max(\mathbf{c})$ are held fixed as $N \to\infty$, so that $m,n \to\infty$.  Greenhill and McKay (GMK)~\cite{greenhill2008asymptotic} improved on this estimate with correction terms thus:
\begin{align}
\begin{split}
\Omega^{\text{GMK}}(\mathbf{r},\mathbf{c}) &= \frac{N!}{\prod_{i = 1}^m r_i! \prod_{j = 1}^n c_j!} \exp\biggl[ \frac{R_{2} C_{2}}{2 N^{2}}+\frac{R_{2} C_{2}}{2 N^{3}}+\frac{R_{3} C_{3}}{3 N^{3}}-\frac{R_{2} C_{2}\left(R_{2}+C_{2}\right)}{4 N^{4}} \\
&\hspace{14em}{} -\frac{R_{2}^{2} C_{3}+R_{3} C_{2}^{2}}{2 N^{4}}+\frac{R_{2}^{2} C_{2}^{2}}{2 N^{5}}+\Ord\biggl(\frac{m^{3} n^{3}}{N^{2}}\biggr) \biggr], 
\end{split}\label{eq:GMKEstimate}
\end{align}
where
\begin{align}
R_k = \sum_{i = 1}^m [r_i]_k\,, \qquad C_k = \sum_{j=1}^n [c_j]_k\,,
\end{align}
and $[x]_k$ is the falling factorial
\begin{align}
[x]_k = x(x-1)\ldots(x - k + 1).
\end{align}
Given the $\Ord\bigl(m^3 n^3/N^2\bigr)$ form of the error term, these estimates are asymptotically correct for $\log \Omega(\mathbf{r},\mathbf{c})$ as $n,m,N \to\infty$ for sufficiently sparse matrices with $mn \sim \textrm{o}(N^{2/3})$.  Note that if we keep only the first term in the exponent of~\eqref{eq:GMKEstimate} we recover the BBK estimate, Eq.~\eqref{eq:BBKEstimate}, so the GMK estimate can be viewed as a correction to the BBK estimate better tailored to the sparse limit.

The numerical performance of both the BBK and GMK estimates is shown in Fig.~\ref{fig:Grid-Random}.  Both perform well in the sparse limit, as one might expect, but are poor in denser regimes.

\section{Maximum-entropy estimates}
\label{maximumEntropy}
Barvinok and Hartigan \cite{barvinok2010maximum} have developed maximum-entropy estimates for a variety of counting problems.  As an application they have given two approximations for counts of contingency tables, a simpler (and faster) Gaussian approximation and a more refined approximation that incorporates a so-called Edgeworth correction.

\subsection{Gaussian maximum-entropy estimate}
\label{ME-GEstimate}
Barvinok and Hartigan~\cite{barvinok2010maximum} give a ``Gaussian'' approximation for~$\Omega(\mathbf{r},\mathbf{c})$, which under quite general conditions on $\mathbf{r}$ and~$\mathbf{c}$ can be shown to return an asymptotically correct value of $\log \Omega(\mathbf{r},\mathbf{c})$ as $N \to\infty$. The approximation takes the form
\begin{align}
\Omega^{\text{G}}(\mathbf{r},\mathbf{c}) &= \frac{e^{g(Z)}}{(2 \pi)^{(m+n-1)/2}\sqrt{\det Q}}.
\label{eq:GEstimate}
\end{align}
Here $g(Z)$ is a scalar function of a matrix~$Z = (z_{ij})$ defined thus:
\begin{align}
g(Z) = \sum_{ij} \left[(z_{ij} + 1) \log(z_{ij}+1) - z_{ij} \log z_{ij}\right].
\label{eq:ME-gX}
\end{align}
The value of~$Z$ is chosen to maximize this function over the same polytope~$P(\mathbf{r},\mathbf{c})$ introduced in Eq.~\eqref{eq:Pdef}, the space of all matrices with non-negative real entries (not necessarily integers) that marginalize to $\mathbf{r}, \mathbf{c}$.  Note that, since $g(Z)$ is concave, there is a unique $Z$ that maximizes it within the polytope.  In practice, the maximum is found numerically with one of the many standard methods for convex optimization.

The quantity~$Q$ in Eq.~\eqref{eq:GEstimate} is an $(m+n-1) \times (m+n-1)$ matrix $Q = (q_{ij})$ whose non-zero elements are
\begin{align}
q_{i,j+m} = q_{i+m,j} &= z_{i j}^{2}+z_{ij} \hspace{3.2em}
  \mbox{for $i=1\ldots m$, $j=1\ldots n$,} \nonumber\\
\label{eq:defsQ}
q_{ii} &= r_{i} + \sum_{j=1}^n z_{ij}^2
  \hspace{2em} \mbox{for $i=1\ldots m$,} \\
q_{j+m,j+m} &= c_{j}+\sum_{i=1}^m z_{ij}^2
  \hspace{2em} \mbox{for $j=1\ldots n-1$.} \nonumber
\end{align}
The computation of the determinant of this matrix in Eq.~\eqref{eq:GEstimate} has time complexity $\Ord((m+n)^3)$ and hence the evaluation of the entire estimate takes at least this long, which makes this method substantially more demanding for large matrices than the linear-time methods of Section~\ref{analytic}.

To understand where the Gaussian estimate comes from, consider a probability distribution $P(X|Z)$ over unrestricted non-negative integer matrices~$X=(x_{ij})$ given a matrix $Z = (z_{ij})$ of real parameters.  Each integer matrix element~$x_{ij}$ is independently drawn from a geometric distribution with expectation~$z_{ij}$, which means the full distribution is
\begin{align}
    P(X|Z) = \prod_{ij} \left(\frac{1}{1 + z_{ij}}\right) \left(\frac{z_{ij}}{1 + z_{ij}}\right)^{x_{ij}}.
\end{align}
The entropy of this probability distribution is equal to the function~$g(Z)$ defined in Eq.~\eqref{eq:ME-gX} and the value of $Z$ is chosen to maximize this entropy over the polytope $Z \in P(\mathbf{r},\mathbf{c})$.

Barvinok and Hartigan now prove a remarkable result, that for this specific choice of~$Z$ the distribution $P(X|Z)$ becomes independent of~$X$ for all $X\in P(\mathbf{r},\mathbf{c})$, taking a constant value equal to
\begin{align}
P(X|Z) = e^{-g(Z)}.
\end{align}
Given that there are, by definition, $\Omega(\mathbf{r},\mathbf{c})$~values of~$X$ inside the polytope, the total probability that $X$ lies in the polytope, and hence that it has the correct margins~$\mathbf{r}$ and~$\mathbf{c}$, is $P\{X \in P(\mathbf{r},\mathbf{c})|Z\} = e^{-g(Z)} \Omega(\mathbf{r},\mathbf{c})$, and hence
\begin{align}
\Omega(\mathbf{r},\mathbf{c}) = e^{g(Z)} P\{X \in P(\mathbf{r},\mathbf{c})|Z\}. \label{eq:omegaME-RC}
\end{align}
Thus, if we can calculate the probability that $X$ has the correct margins we can calculate~$\Omega(\mathbf{r},\mathbf{c})$.  To do this, we observe that the polytope $P(\mathbf{r},\mathbf{c})$ is defined by a set of linear constraints with the general form $AX = b$, where $A$ is an $(m+n-1) \times mn$ matrix, $b$~is an $(m+n-1)$-vector, and $X$ is now represented in ``unrolled'' form as an $mn$-element vector rather than an $m\times n$ matrix.  We then consider the $(m+n-1)$-dimensional random variable $Y = AX$.  This transformed variable satisfies $Y = AX = b$ on $P(\mathbf{r},\mathbf{c})$ and hence $P\{X \in P(\mathbf{r},\mathbf{c})|Z\} = P\{Y = b|Z\}$.

Finally, since the entries of $X$ are independent random variables we expect the distribution of $Y$ to be asymptotically Gaussian by the local central limit theorem.  This allows us to approximate the distribution with a Gaussian, and, matching the covariances of this Gaussian with the true covariances of~$Y$ which are captured in the matrix~$Q$ of Eq.~\eqref{eq:defsQ}, we can estimate $P\{Y = b|Z\}$ and hence the value of $\Omega(\mathbf{r},\mathbf{c})$, yielding the Gaussian maximum-entropy estimate of Eq.~\eqref{eq:omegaME-RC}.

\subsection{Edgeworth correction}
\label{ME-EEstimate}
Building on the Gaussian approximation, Barvinok and Hartigan~\cite{barvinok2012matrices} have given a further improved approximation for $\Omega(\mathbf{r},\mathbf{c})$ by employing a so-called Edgeworth correction. This takes the form
\begin{align}
\Omega^{\text{E}}(\mathbf{r},\mathbf{c}) &= \frac{e^{g(Z)}}{(2 \pi)^{(m+n-1)/2}\sqrt{\det Q}}\exp\left( - \frac{\mu}{2} + \nu\right),
\label{eq:EEstimate}
\end{align}
where $\mu$ and $\nu$ are defined below. Barvinok and Hartigan show that under some mild conditions on the growth of the margins~$\mathbf{r}$ and~$\mathbf{c}$, this gives an asymptotically correct estimate of $\Omega(\mathbf{r},\mathbf{c})$ as $N \rightarrow \infty$. 

To specify the values of~$\mu$ and~$\nu$ in Eq.~\eqref{eq:EEstimate} a few more definitions are needed.  First, we define a quadratic form $q:\mathbb{R}^{m+n-1} \to \mathbb{R}$ by
\begin{align}
q(x) = \tfrac{1}{2} x^T Q x,
\label{eq:quadratic}
\end{align}
where $Q$ is the matrix defined in Eq.~\eqref{eq:defsQ}.  We also define two functions $f,h: \mathbb{R}^{m + n - 1} \to \mathbb{R}$ on the variables $(u_1,\ldots,u_m,t_1,\ldots,t_{n-1}) \in \mathbb{R}^{m+n-1}$ thus:
\begin{align}
f(u, t) &= \frac{1}{6} \sum_{\substack{1 \leq i \leq m \\ 1 \leq j \leq n - 1}} z_{i j}\left(z_{i j}+1\right)\left(2 z_{i j}+1\right)\left(u_{i}+t_{j}\right)^3, \\
h(u, t) &= \frac{1}{24} \sum_{\substack{1 \leq i \leq m \\ 1 \leq j \leq n - 1}} z_{i j}\left(z_{i j}+1\right)\left(6 z_{i j}^{2}+6 z_{i j}+1\right)\left(u_{i}+t_{j}\right)^4.
\end{align}
The Edgeworth correction terms are then given by
\begin{align}
\mu = \mathbf{E}(f^2), \qquad \nu = \mathbf{E}(h),
\end{align}
where the expectations are taken over the Gaussian probability density on $\mathbb{R}^{m+n-1}$ proportional to~$e^{-q}$.

Barvinok and Hartigan also note that, given the definition~\eqref{eq:quadratic}, $Q^{-1}$~is the covariance matrix of the $u_i$ and $t_j$ under the distribution~$e^{-q}$.  This distribution $e^{-q}$ is symmetric under $(u_i,t_j) \to (-u_i,-t_j)$ so that $\mathbf{E}(u_i) = \mathbf{E}(t_j) = 0$ and hence \begin{align}
    \mathbf{E}(u_it_j) = (Q^{-1})_{i (j+m)}\,. \label{eq:utCovariances}
\end{align}
The values of $\mu$ and $\nu$ can then be evaluated using Wick contractions for correlators of Gaussian random variables to express the expectations in terms of covariances given by Eq.~(\ref{eq:utCovariances}).  Specifically, one uses
\begin{align}
\mathbf{E}\bigl[(u_i + t_j)^4\bigr]
  = 3 \bigl[ \mathbf{E}(u_i^2) + 2 \mathbf{E}(u_i t_j)
    + \mathbf{E}(t_j^2)\bigr]^2
\end{align}
and
\begin{align}
\mathbf{E}\bigl[(u_{i_1} + t_{j_1})^3 (u_{i_2} + t_{j_2})^3\bigr]
   &= 3 \bigl[ \mathbf{E}(u_{i_1} u_{i_2}) + \mathbf{E}(u_{i_1} t_{j_2})
    + \mathbf{E}(u_{i_2} t_{j_1}) + \mathbf{E}(t_{i_1} t_{j_2}) \bigr] \nonumber\\
   &\qquad{}\times \bigl[ \mathbf{E}(u_{i_1} u_{i_2})
   + 2 \bigl( \mathbf{E}(u_{i_1}t_{j_2}) + \mathbf{E}(u_{i_2} t_{j_1})
   + \mathbf{E}(t_{j_1}t_{j_2}) \bigr)^2 \nonumber\\
   &\hspace{4em}{} + 3 \bigl( \mathbf{E}(u_{i_1}^2) + 2 \mathbf{E}(u_{i_1} t_{j_1})
   + \mathbf{E}(t_{j_1}^2) \bigr)
   \bigl( \mathbf{E}(u_{i_2}^2) + 2 \mathbf{E}(u_{i_2} t_{j_2})
   + \mathbf{E}(t_{j_2}^2) \bigr)\bigr].
\end{align}
Note that evaluating $\mu$ requires a sum over all possible $i_1,i_2 = 1\dots m$ and $j_1,j_2 = 1\ldots n-1$ so the complexity of the calculation is~$\Ord(m^2n^2)$, making the computational burden higher than for just the Gaussian estimate.  In practice, the running time of either of the maximum-likelihood estimates is not significant for small matrices: our implementations of both run in under a second for $m, n \lesssim 32$.  On the other hand, very large matrices of size $m, n \gtrsim 512$ can take well over an hour, and running time can also be an issue when one needs estimates for a large number of smaller matrices.  For cases where running time is a concern, Section~\ref{app:validationDetails} of the appendices gives our recommendations for various parameter values.

\section{Sequential importance sampling}
\label{SIS}
Sequential importance sampling (SIS) is a computational technique that in the present case can be used either to sample from the set of non-negative integer matrices~$A(\mathbf{r},\mathbf{c})$~\cite{chen2005sequential,harrison2013importance} or to find the size~$\Omega(\mathbf{r},\mathbf{c})$ of the set.  In this section, we review the standard SIS approach and show how it can be refined by exploiting our new linear-time estimate, giving substantially improved performance.

The main ingredient of SIS is a ``trial distribution''~$q(X)$ over matrices~$X$ that is nonzero if and only if $X\in A(\mathbf{r},\mathbf{c})$.  If we can sample matrices from this distribution then we have
\begin{align}
\mathbf{E}_q \left[\frac{1}{q(X)}\right] = \sum_{X \in A(\mathbf{r},\mathbf{c})} q(X)\frac{1}{q(X)} = |A(\mathbf{r},\mathbf{c})| = \Omega(\mathbf{r},\mathbf{c}).
\end{align}
Thus if we can draw $N$ matrices $X^{(1)}\ldots X^{(N)}$ from~$q(X)$ we can estimate $\Omega(\mathbf{r},\mathbf{c})$ as
\begin{align}
\widehat{\Omega}(\mathbf{r},\mathbf{c}) = \frac{1}{N} \sum_{i = 1}^N \frac{1}{q(X^{(i)})}\,,
\label{eq:SIS1}
\end{align}
and the accompanying statistical error can be estimated in the conventional manner.

Though this process works in principle regardless of the form of~$q(X)$, it becomes more efficient the closer the distribution is to the uniform distribution over~$A(\mathbf{r},\mathbf{c})$, since the value of the sum in~\eqref{eq:SIS1} is dominated by the states with the smallest~$q(X)$, which are unlikely to be sampled when $q(X)$ is highly nonuniform.  The key to making the SIS method work well lies in finding a~$q(X)$ that is sufficiently close to the uniform distribution while still being straightforward to work with.  The latter condition can be difficult to satisfy.  We can trivially choose $q(X)$ to be exactly uniform by setting its value to a constant, but in that case the constant is $q(X) = 1/\Omega(\mathbf{r},\mathbf{c})$, so calculating the required value of $q(X)$ would be exactly as hard as calculating $\Omega(\mathbf{r},\mathbf{c})$ in the first place.

SIS gets around these difficulties by sampling the matrix~$X$ one column at a time.  (This is the ``sequential'' part of sequential importance sampling.)  The goal is to sample values~$X_1$ of the first column of~$X$ with probabilities as close as possible to the probability with which they appear under the uniform distribution, which can be written as
\begin{align}
p(X_1) = \frac{\Omega(\mathbf{r}',\mathbf{c}')}{\Omega(\mathbf{r},\mathbf{c})},
\label{eq:pX1}
\end{align}
where $\mathbf{r}'$ and $\mathbf{c}'$ denote the row and column sums of the matrix after the first column is removed.

After the first column is sampled we repeat the process and sample values of the second column, then the third, and so forth until one has a sample of the entire matrix.  If at each step the exact probabilities $p(X_i)$ in Eq.~(\ref{eq:pX1}) are used, this process will sample the matrices $X \in A(\mathbf{r},\mathbf{c})$ exactly uniformly, and indeed this is the approach taken by some methods~\cite{miller2013exact}, although these methods are computationally costly and moreover require us to calculate $\Omega(\mathbf{r},\mathbf{c})$ exactly and hence they are not suitable for calculating $\Omega(\mathbf{r},\mathbf{c})$ itself.

For most purposes a better approach is to approximate the exact distribution~$p(X_1)$ of Eq.~\eqref{eq:pX1} with some other distribution~$q(X_1)$ that is easier to compute at the expense of modestly nonuniform sampling.  Despite this nonuniformity, we can still derive an unbiased estimate for $\Omega(\mathbf{r},\mathbf{c})$ using Eq.~\eqref{eq:SIS1}.  In choosing a value for~$q(X_1)$ the various estimates for $\Omega(\mathbf{r},\mathbf{c})$ in Section~\ref{analytic} provide an elegant route forward and specifically, given its good performance on test cases, we propose using our ``effective columns'' estimate $\Omega^{\text{EC}}(\mathbf{r},\mathbf{c})$ of Eq.~\eqref{eq:ECEstimate} to define a distribution over the column $X_1 = (x_{i1})$ thus:
 \begin{align}
q(X_1) = \frac{\Omega^{\text{EC}}(\mathbf{r}',\mathbf{c}')}{\Omega^{\text{EC}}(\mathbf{r},\mathbf{c})}
    \propto  \prod_{i=1}^m \binom{r_i - x_{i1} + \alpha_{\mathbf{c}'} - 1}{ \alpha_{\mathbf{c}'}-1}
    \mathbf{1}_{\sum_i x_{i1} = c_1} \mathbf{1}_{0 \le x_{i1} \le r_i}\,. \label{eq:qX1}
\end{align} 
This expression combines our combinatorial estimate with hard constraints that impose the correct column margins $\sum_{i} x_{i1} = c_1$ and prevent any entry from surpassing the value of the remaining row sums $0 \le x_{i1} \le r_i$.

As described by Harrison and Miller~\cite{harrison2013importance}, it is possible to sample the column~$X_1$ from a distribution of the form~\eqref{eq:qX1} in time~$\Ord(m c_1^2)$.  The full SIS method samples each of the $n$ columns in turn for a total time complexity of roughly $\Ord(N^2m/n)$ per iteration.  The performance can be improved by a numerical factor (but not in overall complexity) by arranging the values of $\mathbf{c}$ in non-increasing order.

\subsection{Results}
The method described above performs well, as shown in Fig.~\ref{fig:SIS-Comparison}.  The leftmost panel, labeled ``EC-based SIS,'' shows results for our method while the other panels show two other methods for comparison.  ``GC-based SIS'' employs a similar approach to ours but with a trial distribution based on the Good-Crook (GC) estimate~\cite{good1976application}, which appears to have the second-best broad performance behind our EC estimate (see Fig.~\ref{fig:Grid-Random}).  We find that the fractional error for the GC-based method is between 10 and 100 times larger than that for the EC-based method.

The third panel in Fig.~\ref{fig:SIS-Comparison}, labeled ``Greedy SIS,'' shows results from the method of Chen, Diaconis, Holmes, and Liu~\cite{chen2005sequential}, which appears to be the best previous SIS method for counting contingency tables.  In this method the entry $x_{11}$ is directly sampled from the distribution
\begin{align}
\mathbf{Pr}(x_{11} = k) \propto \min(r_2,c_1 - k) +\max(0,c_1 + r_1 + r_2 - N - k) + 1,
\end{align}
and similarly for each remaining entry of~$X$.  This approach gives faster sampling than ours, but at the expense of a more non-uniform~$q(X)$.   The trade-off turns out not to be beneficial.  Even though the greedy method can perform 100 or more times as many iterations as the EC-based method in a comparable amount of time, it nonetheless converges more slowly to the answer and overall accuracy suffers, as shown in Fig.~\ref{fig:SIS-Comparison}. 

Based on these results we have choosen to use the EC-based SIS technique to compute ground-truth estimates of $\Omega(\mathbf{r},\mathbf{c})$ in our work.  We emphasize that this does not in any way bias the outcome of our benchmarking comparisons in Fig.~\ref{fig:Grid-Random} in favor of the EC estimate.  All SIS methods, regardless of their choice of trial distribution, give unbiased estimates; the choice of an EC-based trial distribution merely improves the rate of convergence of those estimates.

\section{Conclusions}
In this paper we have studied the problem of estimating the number~$\Omega(\mathbf{r},\mathbf{c})$ of non-negative integer matrices with given row and column sums, which arises for example in statistical and information theoretic calculations involving contingency tables.  There is no known exact expression for~$\Omega(\mathbf{r},\mathbf{c})$, but a variety of methods for approximating it have been proposed in the past.  We have presented two new methods that improve upon these previous approaches.  First, we have proposed a closed-form approximation based on a concept of \defn{effective columns}, which can be evaluated in time linear in the number $m+n$ of rows plus columns of the matrix and returns results of accuracy similar to or better than other linear-time estimates in the extensive benchmark tests presented here.  Second, the same effective-columns approximation is also used to derive a new sequential importance sampling (SIS) algorithm for estimating~$\Omega(\mathbf{r},\mathbf{c})$ with significantly faster convergence than previous SIS methods, resulting in numerical estimates as much as 100 times more accurate in comparable running times.

\section{Acknowledgments}
This work was supported in part by the US National Science Foundation under grant DMS--2005899 and by computational resources provided by the Advanced Research Computing initiative at the University of Michigan.  See \verb|https://github.com/maxjerdee/contingency_count| for code implementing the methods described here.

\section*{Appendices}

\appendix 
\section{Validity of the ``effective columns'' estimate}
\label{app:ECdetails}
The value of the parameter~$\alphac$ in our estimate is given by Eq.~\eqref{eq:alphaC} to be
\begin{align}
\alphac = \frac{N^2 - N + (N^2 - c^2)/m}{c^2 - N}.
\label{eq:alphaApp}
\end{align}
At first sight this expression appears potentially problematic, since the denominator could become zero or negative.  In fact it cannot be negative because
\begin{align}
c^2 - N = \sum_{j=1}^n c_j^2 - \sum_{j=1}^n c_j = \sum_{j=1}^n c_j(c_j-1) \ge 0.
\end{align}
The value could however be zero if $c_j$ is either zero or one for all~$j$, and this would cause $\alphac$ to diverge.  In practice we can ignore columns with $c_j=0$ since these have no effect on the number of matrices~$\Omega(\mathbf{r},\mathbf{c})$, so let us assume that all such columns have been removed.  What then happens if all remaining columns have~$c_j=1$?  In this case it turns out that the limit $\alphac\to\infty$ of the estimate $\Omega^{\text{EC}}(\mathbf{r},\mathbf{c})$ in fact gives the correct result, as we now show.

If all columns have $c_j=1$ then all elements in a column are zero except for a single~1.  The constraint on the row sums then demands that $r_i$ out of the $n$ columns have their 1 in row~$i$ for all~$i=1\ldots m$.  The number of possible arrangements satisfying this requirement is
\begin{align}
\Omega(\mathbf{r},\mathbf{c} = (1,\ldots,1)) &= \frac{n!}{\prod_{i=1}^m r_i!}. \label{eq:exactOmega1s}
\end{align}
We now show that the $\alphac \to \infty$ limit of our estimate Eq.~\eqref{eq:ECEstimate} for this situation gives this exact solution.

Recall that our estimate is given by
\begin{align}
\Omega^{\text{EC}}(\mathbf{r},\mathbf{c})
    = \binom{N+m \alphac - 1}{m \alphac - 1}^{-1} \prod_{i=1}^m \binom{r_i+\alphac - 1}{\alphac - 1} \prod_{j=1}^n \binom{c_j+m-1}{m - 1}.
\label{eq:ECEstimateAgain}
\end{align}
Noting that when $c_j=1$ for all $j$ we have $N=n$, we write
\begin{align}
\binom{N+m \alphac - 1}{m \alphac - 1} = \frac{\Gamma(n + m \alphac)}{\Gamma(m \alphac)\Gamma(n + 1)}, \qquad
\binom{r_i + \alphac - 1}{\alphac - 1} = \frac{\Gamma(r_i + \alphac)}{\Gamma(\alphac)\Gamma(r_i + 1)},
\end{align}
and apply Stirling's approximation in the form
\begin{align}
\Gamma(z) = \sqrt{\frac{2\pi}{z}} \left(\frac{z}{e}\right)^z\left(1 + \Ord(z^{-1})\right),
\end{align}
which in the limit of large~$\alphac$ gives
\begin{equation}
    \binom{n+m \alphac - 1}{m \alphac - 1} = \frac{(m \alphac)^n}{\Gamma(n+1)}\left(1 + \Ord(\alphac^{-1})\right),\qquad
    \binom{r_i + \alphac - 1}{\alphac - 1} = \frac{\alphac^{r_i}}{\Gamma(r_i + 1)}(1 + \Ord(\alphac^{-1})).
\end{equation}

Then our estimate, Eq.~\eqref{eq:ECEstimateAgain}, is
\begin{align}
\binom{n+m \alphac - 1}{m \alphac - 1}^{-1} \prod_{i=1}^m \binom{r_i + \alphac - 1}{ \alphac - 1} \prod_{j=1}^n \binom{1+m-1}{m - 1}
  &= \frac{\Gamma(n + 1)}{(m \alphac)^n} \prod_{i=1}^m \frac{\alphac^{r_i}}{\Gamma(r_i + 1)} \prod_{j=1}^n m \left(1 + \Ord(\alphac^{-1})\right) \nonumber\\
  &= \frac{n!}{\prod_{i=1}^m r_i!}\left(1 + \Ord(\alphac^{-1})\right). 
\end{align}
So the $\alphac \to \infty$ limit indeed recovers the correct result. 

This is a nice property of our estimate.  In a practical implementation we can recognize the case $\mathbf{c} = (1,\ldots,1)$ and either return the exact result, Eq.~\eqref{eq:exactOmega1s}, or simply evaluate the usual estimate at a large value of~$\alphac$.  The latter prescription is also equivalent to writing
\begin{align}
\alphac = \frac{N^2 - N + (N^2 - c^2)/m}{c^2 - N + \epsilon}
\end{align}
for $\epsilon$ small and positive.

\section{Numerical calculations}
\label{app:numericalDetails}
In this appendix we give some technical details of the numerical tests reported in Section~\ref{summary}.

\subsection{Generation of test cases}
\label{app:testCaseGeneration}
The process by which the test values of $\mathbf{r}, \mathbf{c}$ are sampled for benchmarking can impact results like those in Fig.~\ref{fig:Grid-Random}.  In this section, we describe the scheme we use, explore the impact of using a different scheme, and examine the effect of changing the shape~$m,n$ of the matrix while keeping the sum~$N$ of all elements fixed.  In all cases we find that our new estimate for~$\Omega(\mathbf{r},\mathbf{c})$ appears to outperform other linear-time methods.

In generating the test cases we choose to sample uniformly over possible margins~$\mathbf{r}$ and~$\mathbf{c}$ that have the required sizes $m,n$ and sum to a given~$N$.  We also require that all $r_i,c_j$ be nonzero, since cases with zeros can be trivially simplified by removing the zeros.  Thus, for example, $\mathbf{r}$~is drawn uniformly from the set of all $m$-element vectors with strictly positive integer entries that sum to~$N$.

There are other possible approaches, however.  One could sample the margins by first generating a matrix, sampled uniformly from the set of non-negative integer $m \times n$ matrices that sum to~$N$, and then take the row and column sums of this matrix to form~$\mathbf{r}$ and~$\mathbf{c}$.  In effect, this process samples the margins~$\mathbf{r}$ and~$\mathbf{c}$ weighted by the number of possible matrices~$\Omega(\mathbf{r},\mathbf{c})$ with those margins.  In practice this yields more uniform margins, particularly for larger and denser matrices, because there are larger numbers of matrices with relatively uniform margins than with non-uniform ones.

\begin{figure}
\centering
\includegraphics[width=\textwidth]{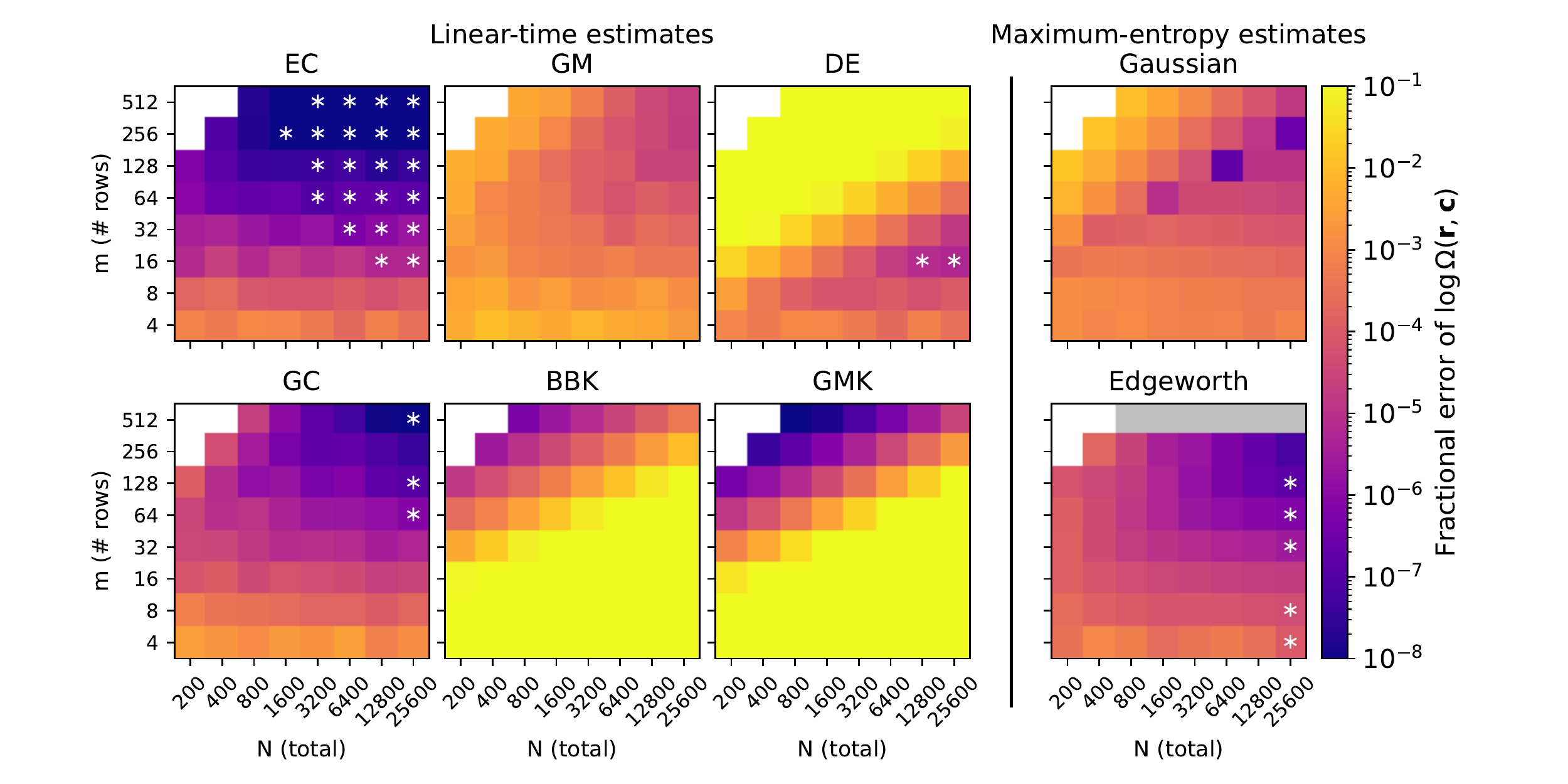}
\caption{Fractional error in various estimates of~$\log \Omega(\mathbf{r},\mathbf{c})$.  As in Fig.~\ref{fig:Grid-Random} the results are for square $m \times m$ matrices of total sum~$N$.  Unlike Fig.~\ref{fig:Grid-Random}, on the other hand, the values of $\mathbf{r},\mathbf{c}$ are the observed margins of uniformly sampled matrices, rather than being uniformly sampled themselves.}
\label{fig:Grid-Uniform}
\end{figure}

Making the margins more uniform typically improves the accuracy of our estimates for~$\Omega(\mathbf{r},\mathbf{c})$, as shown in Fig.~\ref{fig:Grid-Uniform}.  Comparing to Fig.~\ref{fig:Grid-Random} we see that all of the estimates generally perform better for the more uniform margins.  Our EC estimate, however, still stands out as performing particularly well and moreover is now competitive with the SIS benchmark and with the maximum-entropy methods for large $N$ and~$m$.  These results suggest that the more uniform margins comprise the ``easy cases'' for approximating~$\Omega(\mathbf{r},\mathbf{c})$ and the more heterogeneous margins of Fig.~\ref{fig:Grid-Random} provide a more stringent test. 

\begin{figure}
\centering
\includegraphics[width=\textwidth]{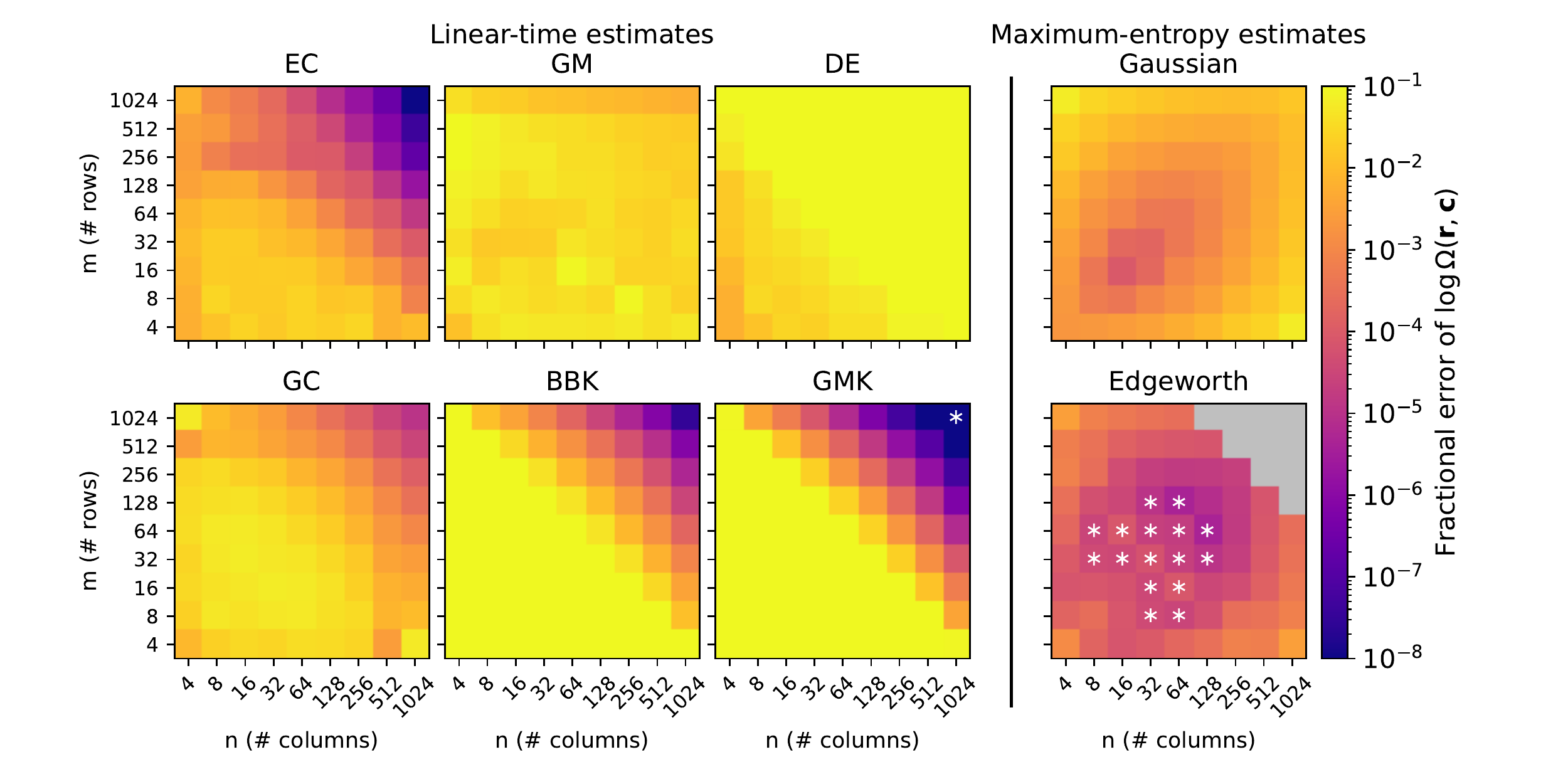}
\caption{Fractional error on various estimates of~$\log \Omega(\mathbf{r},\mathbf{c})$.  In these tests the totals of all matrices are fixed at $N = 1600$ while the number of rows and columns is varied.  Each data point is averaged over ten sets of margins~$\mathbf{r},\mathbf{c}$ drawn uniformly at random.  We observe that the performance of the linear-time estimates depends chiefly on the product~$mn$, while the maximum-entropy estimates struggle with highly oblong matrices.}
\label{fig:Grid-Shape}
\end{figure}

In Fig.~\ref{fig:Grid-Random} we also chose to consider only square $m \times m$ matrices, since performance seems to be driven primarily by the value of $N$ and the product of the dimensions~$mn$ only.  Figure~\ref{fig:Grid-Shape} offers some evidence for this claim.  In this figure we show the results of tests in which $m$ and $n$ are varied while keeping $N$ fixed at a value of 1600, and we see that most of the performance is indeed explained by the combination~$mn$---constant $mn$ in this figure corresponds to diagonal lines from top-left to bottom-right.  Some of the linear-time estimates (EC, GM, and DE), however, do show some mild asymmetry, and the maximum-entropy methods have difficulty when $m$ and $n$ are very different. These patterns are also observed for other choices of~$N$.

\subsection{Benchmarking}
\label{app:validationDetails}
Benchmarking of approximation methods requires us to compute accurate ground-truth estimates of~$\Omega(\mathbf{r},\mathbf{c})$ for comparison.  In this section we describe various methods for doing this, and in particular address the following question: if you have one hour of computation time (on common hardware \textit{circa}~2022) to get the highest quality estimate of $\Omega(\mathbf{r},\mathbf{c})$, what method should you use?  Under these conditions, linear-time estimates never give the best answer (although in applications where speed is important, such as when estimating $\Omega(\mathbf{r},\mathbf{c})$ for a large number of small matrices, linear-time methods may be the best).

\begin{figure}[t]
\centering
\includegraphics[width=0.45\textwidth]{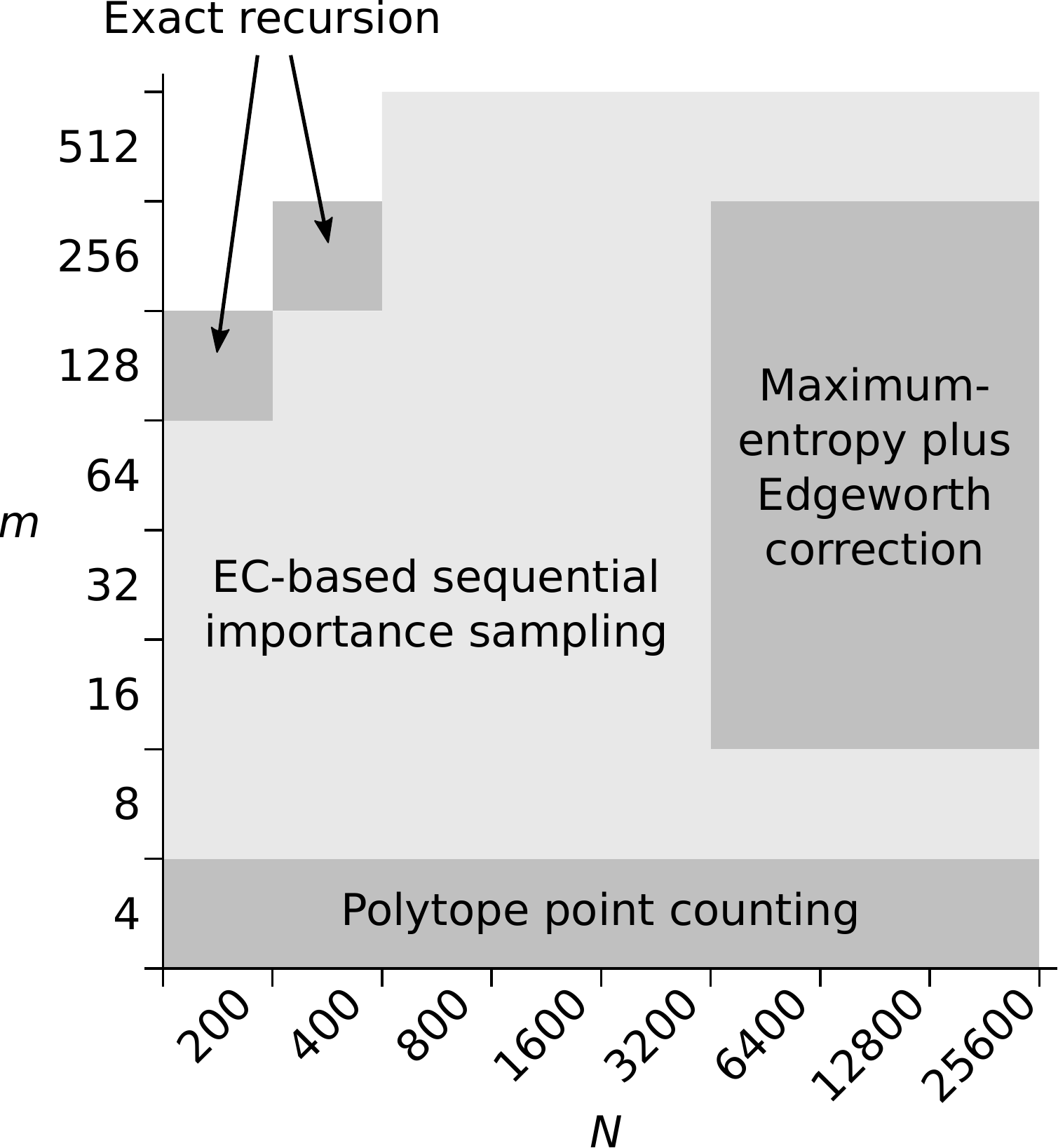}
\caption{Schematic of which method gives the best performance for estimating~$\Omega(\mathbf{r},\mathbf{c})$ within one hour of runtime.  The exact methods shown are Barvinok's polytope algorithm~\cite{barvinok1994polynomial, baldoni2014user} and the recursion-based approach of Harrison and Miller~\cite{miller2013exact}.  For many large $N$ cases the Edgeworth-corrected maximum-entropy estimate is the winner, while for all others the EC-based sequential importance sampling approach of this paper is the method of choice.  The white region top left represents invalid parameter choices for which there are no matrices with the given margins.  This diagram also represents the choice of method used for generating test configurations $m,N$ to validate the SIS results in Fig.~\ref{fig:SIS-Comparison}. }
\label{fig:Validation-Regions}
\end{figure}

Figure~\ref{fig:Validation-Regions} summarizes our results for the best method to use as a function of $m$ and~$N$.  In certain regimes exact solutions are available.  Barvinok's algorithm for counting integer points in convex polytopes~\cite{barvinok1994polynomial,barvinok1999algorithmic} can be applied to give an exact algorithm with running time polynomial in~$N$, which we implement using the \verb|count| function from the \verb|lattE| software package~\cite{baldoni2014user}.  This allows very large values of~$N$ to be probed, but the complexity grows quickly in~$m$ so this method is limited to $m \lesssim 6$ on current hardware.

In sparse situations with bounded margins $\mathbf{r}$ and~$\mathbf{c}$ we can compute $\Omega(\mathbf{r},\mathbf{c})$ exactly using recursion-based methods.  Harrison and Miller~\cite{miller2013exact} have given an implementation of this approach which exploits repeated entries in the margins to improve running time.  While not shown in Fig.~\ref{fig:Validation-Regions}, this method can also be used for most cases where $N \lesssim 100$.

For all other cases, we use approximate ground-truth estimates of~$\Omega(\mathbf{r},\mathbf{c})$ computed using sequential importance sampling (SIS), and among the various SIS methods the EC-based method of this paper (Section~\ref{SIS}) performs the best as shown in Fig.~\ref{fig:SIS-Comparison}.  In principle, the Edgeworth-corrected maximum-entropy method of Barvinok and Hartigan~\cite{barvinok2012matrices} (Section~\ref{ME-EEstimate}) outperforms SIS in certain regimes as can be seen in Fig.~\ref{fig:Grid-Random}.  This is not useful for benchmarking, however, since this is one of the approximations we are trying to evaluate, but in a more general setting where one simply wanted to make the best estimate of~$\Omega(\mathbf{r},\mathbf{c})$ in the time allotted the maximum-entropy method could be useful in certain regimes.  Where exactly this occurs is somewhat arbitrary, but this informs the boundary in Fig.~\ref{fig:Validation-Regions}.

\section{Counting 0-1 matrices}
\label{app:ZeroOneTables}
A parallel problem to that of counting non-negative integer matrices is that of counting matrices whose elements take only the values zero and one, with row and columns sums once again fixed at given values~$\mathbf{r}$ and~$\mathbf{c}$.  This problem is important in its own right and has been the subject of considerable work.  The methods of this paper can be applied to the case of 0-1 matrices, but we have not emphasized this approach because in practice it does not improve upon existing methods.  Nonetheless, for the sake of completeness, we describe the approach in this appendix.

\subsection{Summary of results}
\begin{figure}
\centering
\includegraphics[width=\textwidth]{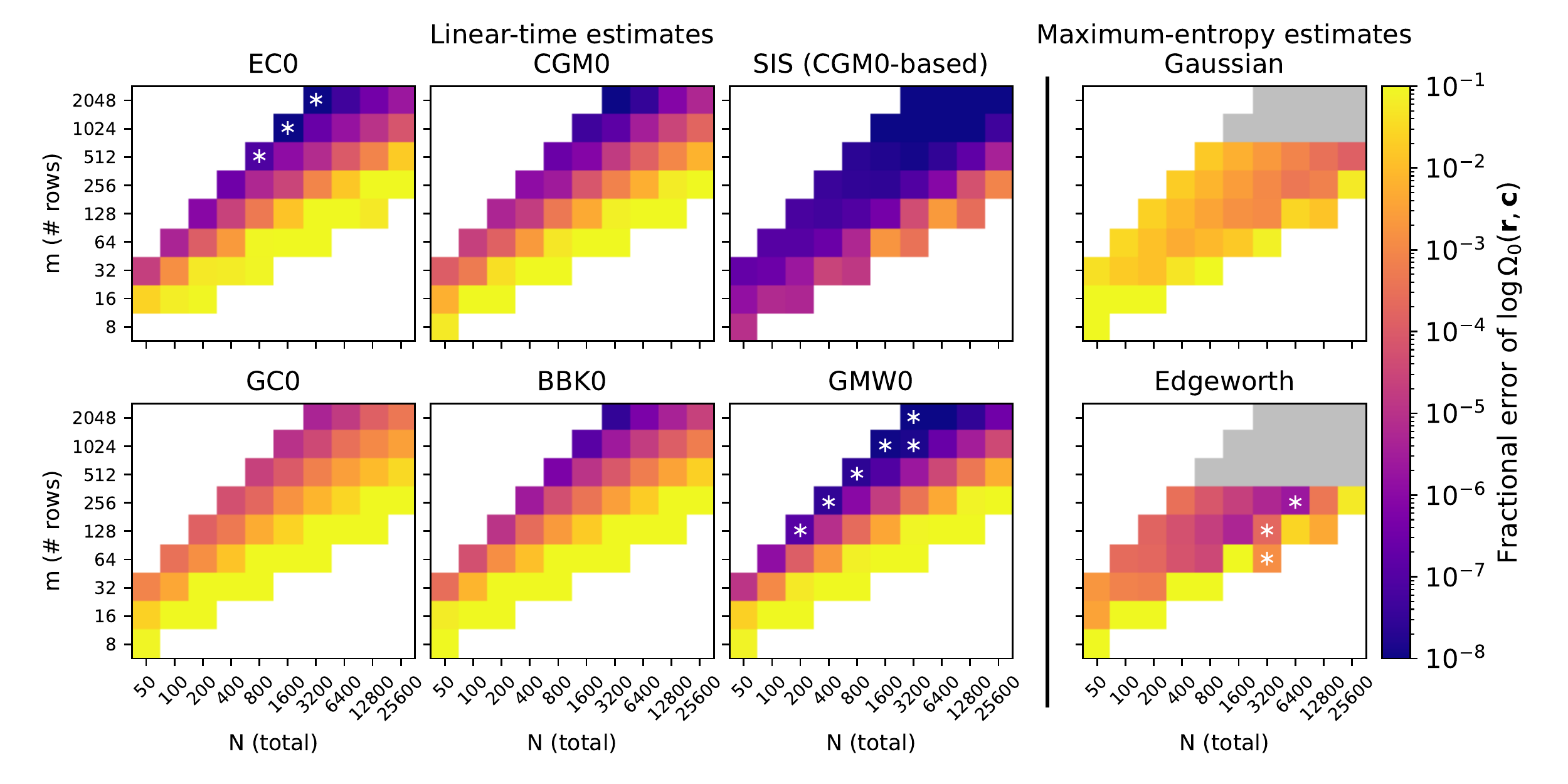}
\caption{Fractional error on various estimates of $\log \Omega_0(\mathbf{r},\mathbf{c})$ for square $m \times m$ matrices that sum to~$N$.  For compactness the estimated errors of the SIS method used for benchmarking are also plotted under the linear-time category.}
\label{fig:Grid0}
\end{figure}

Let $\Omega_0(\mathbf{r},\mathbf{c})$ be the number of 0-1 matrices with margins~$\mathbf{r},\mathbf{c}$.  Figure~\ref{fig:Grid0} summarizes the performance of various estimates of $\log \Omega_0(\mathbf{r},\mathbf{c})$ for square $m \times m$ matrices that sum to~$N$. In the linear-time category we consider five estimates.  The first, an analogue of the effective columns estimate for the case of non-negative integer matrices, is given by
\begin{align}
\Omega_0^{\text{EC}}(\mathbf{r},\mathbf{c}) &= \left|\binom{m \alphacz}{N}^{-1} \prod_{i=1}^m \binom{\alpha_\mathbf{c}^{(0)}}{r_i} \prod_{j=1}^n \binom{m}{c_j}\right|,
\end{align}
where
\begin{align}
\alphacz = \frac{N^2 - N - (N^2 - c^2)/m}{c^2 - N}.
\end{align}
While this estimate performs fairly well, it does not reliably outperform the other linear-time estimates.  In general, all of the linear-time methods struggle with dense matrices. 

To generate Fig.~\ref{fig:Grid0}, for each combination of parameters $N,m$, ten margin pairs $\mathbf{r},\mathbf{c}$ were drawn uniformly from the set of all values that correspond to at least one 0-1 matrix (i.e.,~uniformly over values satisfying the \defn{Gale-Ryser condition}~\cite{gale1957theorem,ryser1957combinatorial}).  The ground truth is computed using sequential importance sampling as described in Section~\ref{SIS} with a trial distribution based on the CGM0 estimate as in~\cite{harrison2013importance}.  The resulting estimated errors on the sequential importance sampling are also shown in Fig.~\ref{fig:Grid0}.

\subsection{Effective columns estimate}
Motivated by the same ``effective columns'' reasoning as in our estimate for the number of contingency tables, we propose the following estimate for the number of 0-1 matrices:
\begin{align}
\Omega_0^{\text{EC}}(\mathbf{r},\mathbf{c}) &= \left|\binom{m \alpha_\mathbf{c}^{(0)}}{N}^{-1} \prod_{i=1}^m \binom{\alpha_\mathbf{c}^{(0)}}{r_i} \prod_{j=1}^n \binom{m}{c_j}\right|,
\label{eq:EC0Estimate}
\end{align}
where
\begin{align}
\alpha_{\mathbf{c}}^{(0)} = \frac{N^2 - N - (N^2 - c^2)/m}{c^2 - N}.
\end{align}
This estimate stands on less certain ground than our estimate of~$\Omega(\mathbf{r},\mathbf{c})$, but the resulting formula appears to be quite accurate so we briefly outline the rationale. 

Let $A_0(\mathbf{c})$ be the set of 0-1 matrices that have column sums~$\mathbf{c}$.  The number of such matrices can be found by independently choosing one column at a time.  For each column~$j$ there are $c_j$ elements equal to 1 and the rest are equal to~0, so there are $\binom{m}{c_i}$ ways to distribute the 1s in the column.  Since the columns are independent we then have
\begin{align}
|A_0(\mathbf{c})| = \prod_{j=1}^n \binom{m}{c_j}.
\end{align}
Given this number, $\Omega(\mathbf{r},\mathbf{c})$~can be estimated as before from a knowledge of the conditional distribution~$\mathbf{Pr}(\mathbf{r}|\mathbf{c})$, and for this we again take inspiration from the unconditional distribution of~$\mathbf{r}$,
\begin{align}
\mathbf{Pr}(\mathbf{r}) = \binom{mn}{N}^{-1} \prod_{i=1}^m \binom{n}{r_i},
\end{align} 
replacing the number of columns~$n$ with an effective number~$\alphacz$:
\begin{align}
\tilde{P}(\mathbf{r}|\alphacz) = \binom{m\alphacz}{N}^{-1} \prod_{i=1}^m \binom{\alphacz}{r_i}.
\label{eq:Pralpha0}
\end{align}
Unlike the case of non-negative integer matrices, where we were left with a well-defined distribution for any $\alphac > 0$, $\tilde{P}(\mathbf{r}|\alphacz)$~is not quite a probability distribution over~$\mathbf{r}$.  Away from the poles, when $\alphacz\notin \bigl\{0,1/m,\ldots,(N-1)/m\bigr\}$, Eq.~\eqref{eq:Pralpha0} is properly normalized
\begin{align}
\sum_{\mathbf{r} | \sum_i r_i = N} \tilde{P}(\mathbf{r}|\alphacz) = 1,
\end{align}
but it is no longer non-negative for all~$\mathbf{r}$: we can have $\tilde{P}(\mathbf{r}|\alphacz) < 0$.  In spite of this we press on and evaluate the ``expectations'' and ``co-variances'' of the $r_i$ weighted by~$\tilde{P}(\mathbf{r}|\alphacz)$:
\begin{align}
\mathbf{E}(r_i) = \frac{N}{m}, \qquad \mathbf{cov}(r_i,r_{k}) = \frac{N(m \alphacz - N)}{m(m \alphacz - 1)} \left(\delta_{ik} - m^{-1}\right).
\end{align}
The true probability density $\mathbf{Pr}(\mathbf{r}|\mathbf{c})$ is again a mixture of independent columns with expectation and covariances
\begin{align}
\mathbf{E}(r_i) = \frac{N}{m}, \qquad \mathbf{cov}(r_i,r_{k}) = \frac{N m - c^2}{m(m-1)} \left(\delta_{ik} - m^{-1}\right).
\end{align}
The choice of the parameter $\alphacz$ such that the moments of $\tilde{P}(\mathbf{r}|\alphacz)$ and the true $\mathbf{Pr}(\mathbf{r}|\mathbf{c})$ match is then
\begin{align}
\alphacz = \frac{N^2 - N - (N^2 - c^2)/m}{c^2 - N},
\end{align}
and our esimate of $\Omega_0(\mathbf{r},\mathbf{c})$ is given by $\tilde{P}(\mathbf{r}|\alphacz)|A_0(\mathbf{c})|$.  In most cases, this expression can be used directly, but in the occasional instances where $\tilde{P}(\mathbf{r}|\alphacz)$ is negative the resulting estimate can be negative as well.  We remedy this problem in an ad-hoc way by taking the absolute value of the result, which yields the estimate Eq.~\eqref{eq:EC0Estimate}.  In spite of this rather arbitrary step the estimate performs reasonably well in the tests shown in Fig.~\ref{fig:Grid0}.

\subsection{Other estimates}
We also consider four other linear-time estimates of $\Omega_0(\mathbf{r},\mathbf{c})$ drawn from the literature, many of which are related to those for the case of general non-negative integer matrices.  Good and Crook~\cite{good1977enumeration} give an estimate which can be understood as our effective columns estimate but with the number of effective columns equal to the number of true columns:
\begin{align}
    \Omega_0^{\text{GC}}(\mathbf{r},\mathbf{c}) &= \binom{m n}{N}^{-1}\prod_{i=1}^m \binom{n}{r_i} \prod_{j=1}^n \log \binom{m}{c_j}. \label{eq:GC0Estimate}
\end{align}
B\'ek\'essy, B\'ek\'essy, and Koml\'os (BBK)~\cite{bekessy1972asymptotic} provide an estimate suited to the sparse regime:
\begin{align}
\Omega_0^{\text{BBK}}(\mathbf{r}, \mathbf{c}) &= \frac{N!}{\prod_{i=1}^m r_i! \prod_{j=1}^m c_j!}\exp\left[-\frac{2}{N^2} \sum_{i=1}^m \binom{r_i}{2} \sum_{j=1}^n \binom{c_j}{2}\right]\,\bigl[1 + \Ord\bigl(\log N/\sqrt{N}\bigr)\bigr], 
\label{eq:BBK0Estimate}
\end{align}
which is improved by Greenhill, McKay, and Wang (GMW)~\cite{greenhill2006asymptotic}:
\begin{align}
\begin{split}
\Omega_0^{\text{GMW}}(\mathbf{r},\mathbf{c}) &= \frac{N!}{\prod_{i = 1}^m r_i! \prod_{j = 1}^n c_j!} \text{exp}\Bigg(-\frac{R_{2} C_{2}}{2 N^{2}}-\frac{R_{2} C_{2}}{2 N^{3}}+\frac{R_{3} C_{3}}{3 N^{3}}-\frac{R_{2} C_{2}\left(R_{2}+C_{2}\right)}{4 N^{4}}\\
&\qquad-\frac{R_{2}^{2} C_{3}+R_{3} C_{2}^{2}}{2 N^{4}}+\frac{R_{2}^{2} C_{2}^{2}}{2 N^{5}}+O\left(\frac{m^{3} n^{3}}{N^{2}}\right)\Bigg). \label{eq:GMW0Estimate}
\end{split} 
\end{align}
Canfield, Greenhill, and McKay~\cite{canfield2008asymptotic} provide an estimate for dense 0-1 matrices that can be understood as a correction to the GC estimate:
\begin{align}
    \Omega_0^{\text{CGM}}(\mathbf{r},\mathbf{c}) = \binom{mn}{N}^{-1} \prod_{i = 1}^m \binom{n}{r_i} \prod_{j = 1}^n \binom{m}{c_j} \exp\left[-\frac{1}{2}\left(1 - \frac{R}{2 A m n}\right)\left(1 - \frac{C}{2 A m n}\right)\right], \label{eq:CGM0Estimate}
\end{align}
where
\begin{align}
R &= \sum_{i=1}^m \left(r_i - \frac{N}{m}\right)^2, \qquad C = \sum_{j=1}^n \left(c_j - \frac{N}{n}\right)^2, \qquad
\lambda = \frac{N}{mn}, \qquad
A = \tfrac{1}{2} \lambda (1 - \lambda).
\end{align}
Canfield~\etal\ show that this is in fact asymptotically correct under certain conditions---loosely when the matrix is relatively square and has density not too close to 0 or 1.  Finally, Barvinok and Hartigan~\cite{barvinok2013number} give maximum-entropy estimates in Gaussian and Edgeworth-corrected varieties analogous to those of Section~\ref{maximumEntropy}.


\begin{thebibliography}{10}
\expandafter\ifx\csname url\endcsname\relax
  \def\url#1{\texttt{#1}}\fi
\expandafter\ifx\csname urlprefix\endcsname\relax\def\urlprefix{URL }\fi

\bibitem{newman2020improved}
M.~E.~J. Newman, G.~T. Cantwell, and J.-G. Young, Improved mutual information
  measure for clustering, classification, and community detection.
  \textit{Physical Review E} \textbf{101}, 042304 (2020).

\bibitem{chen2005sequential}
Y.~Chen, P.~Diaconis, S.~P. Holmes, and J.~S. Liu, Sequential {M}onte {C}arlo
  methods for statistical analysis of tables. \textit{Journal of the American
  Statistical Association} \textbf{100}, 109--120 (2005).

\bibitem{harrison2013importance}
M.~T. Harrison and J.~W. Miller, Importance sampling for weighted binary random
  matrices with specified margins. \textit{arXiv preprint arXiv:1301.3928}
  (2013).

\bibitem{dyer1997sampling}
M.~Dyer, R.~Kannan, and J.~Mount, Sampling contingency tables. \textit{Random
  Structures \& Algorithms} \textbf{10}, 487--506 (1997).

\bibitem{barvinok1994polynomial}
A.~I. Barvinok, A polynomial time algorithm for counting integral points in
  polyhedra when the dimension is fixed. \textit{Mathematics of Operations
  Research} \textbf{19}, 769--779 (1994).

\bibitem{miller2013exact}
J.~W. Miller and M.~T. Harrison, Exact sampling and counting for fixed-margin
  matrices. \textit{The Annals of Statistics} \textbf{41}, 1569--1592 (2013).

\bibitem{good1976application}
I.~J. Good, On the application of symmetric {D}irichlet distributions and their
  mixtures to contingency tables. \textit{The Annals of Statistics} \textbf{4},
  1159--1189 (1976).

\bibitem{good1977enumeration}
I.~J. Good and J.~F. Crook, The enumeration of arrays and a generalization
  related to contingency tables. \textit{Discrete Mathematics} \textbf{19},
  23--45 (1977).

\bibitem{diaconis1985testing}
P.~Diaconis and B.~Efron, Testing for independence in a two-way table: New
  interpretations of the chi-square statistic. \textit{The Annals of
  Statistics} \textbf{13}, 845--874 (1985).

\bibitem{gail1977counting}
M.~Gail and N.~Mantel, Counting the number of $r\times c$ contingency tables
  with fixed margins. \textit{Journal of the American Statistical Association}
  \textbf{72}, 859--862 (1977).

\bibitem{bekessy1972asymptotic}
A.~B{\'e}k{\'e}ssy, Asymptotic enumeration of regular matrices. \textit{Studia
  Scientiarum Mathematicarum Hungarica} \textbf{7}, 343--353 (1972).

\bibitem{greenhill2008asymptotic}
C.~Greenhill and B.~D. McKay, Asymptotic enumeration of sparse nonnegative
  integer matrices with specified row and column sums. \textit{Advances in
  Applied Mathematics} \textbf{41}, 459--481 (2008).

\bibitem{barvinok2012matrices}
A.~Barvinok, Matrices with prescribed row and column sums. \textit{Linear
  Algebra and its Applications} \textbf{436}, 820--844 (2012).

\bibitem{holmes1996uniform}
R.~B. Holmes and L.~K. Jones, On uniform generation of two-way tables with
  fixed margins and the conditional volume test of {D}iaconis and {E}fron.
  \textit{The Annals of Statistics} \textbf{24}, 64--68 (1996).

\bibitem{rodney2010asymptotic}
E.~R. Canfield and B.~D. McKay, Asymptotic enumeration of integer matrices with
  large equal row and column sums. \textit{Combinatorica} \textbf{30}, 655--680
  (2010).

\bibitem{barvinok2010maximum}
A.~Barvinok and J.~A. Hartigan, Maximum entropy {G}aussian approximations for
  the number of integer points and volumes of polytopes. \textit{Advances in
  Applied Mathematics} \textbf{45}, 252--289 (2010).

\bibitem{barvinok1999algorithmic}
A.~Barvinok and J.~E. Pommersheim, An algorithmic theory of lattice points in
  polyhedra. \textit{New Perspectives in Algebraic Combinatorics} \textbf{38},
  91--147 (1999).

\bibitem{baldoni2014user}
V.~Baldoni, N.~Berline, J.~A. De~Loera, B.~Dutra, M.~K{\"o}ppe, S.~Moreinis,
  G.~Pinto, M.~Vergne, and J.~Wu, A user’s guide for {LattE} integrale
  v1.7.2. \textit{Optimization} \textbf{22} (2014).

\bibitem{gale1957theorem}
D.~Gale, A theorem on flows in networks. \textit{Pacific Journal of
  Mathematics} \textbf{7}, 1073--1082 (1957).

\bibitem{ryser1957combinatorial}
H.~J. Ryser, Combinatorial properties of matrices of zeros and ones.
  \textit{Canadian Journal of Mathematics} \textbf{9}, 371--377 (1957).

\bibitem{greenhill2006asymptotic}
C.~Greenhill, B.~D. McKay, and X.~Wang, Asymptotic enumeration of sparse 0--1
  matrices with irregular row and column sums. \textit{Journal of Combinatorial
  Theory, Series A} \textbf{113}, 291--324 (2006).

\bibitem{canfield2008asymptotic}
E.~R. Canfield, C.~Greenhill, and B.~D. McKay, Asymptotic enumeration of dense
  0--1 matrices with specified line sums. \textit{Journal of Combinatorial
  Theory, Series A} \textbf{115}, 32--66 (2008).

\bibitem{barvinok2013number}
A.~Barvinok and J.~A. Hartigan, The number of graphs and a random graph with a
  given degree sequence. \textit{Random Structures \& Algorithms} \textbf{42},
  301--348 (2013).

\end{thebibliography}
\end{document}